\documentclass[11pt]{article}
\renewcommand{\theequation}{\thesection.\arabic{equation}}

\usepackage{amsfonts}
\usepackage{amsbsy}
\usepackage[dvips]{graphicx
}
\newcommand{\BE}{\begin{equation}}
\newcommand{\EE}{\end{equation}}
\newcommand{\BA}{\begin{eqnarray}}
\newcommand{\EA}{\end{eqnarray}}
\newcommand{\half}{{\scriptstyle{\frac{1}{2}}}}

\newcommand{\quarter}{{\scriptstyle{\frac{1}{4}}}}
\newcommand{\Pp}{{\scriptstyle{{\rm P}}}}
\newcommand{\effch}{{\scriptscriptstyle{\rm FAC}}}
\newcommand{\msbar}{\overline{\rm MS}}
\newcommand{\smmsbar}{{\scriptscriptstyle{\overline{\rm MS}}}}
\newcommand{\kplus}{(k+1)^{th}}

\begin{document}

\begin{titlepage}

\vspace*{25mm}
\begin{center}
               {\Large\bf Optimization of QCD Perturbation Theory:} \\
\vspace*{4mm}
  {\Large\bf Results for $R_{e^+e^-}$ at fourth order}
\vspace{19mm}\\
{\Large P. M. Stevenson}
\vspace{7mm}\\
{\large\it
T.W. Bonner Laboratory, Department of Physics and Astronomy,\\
Rice University, Houston, TX 77251, USA}
\vspace{35mm}\\
{\bf Abstract:}
 
\end{center}

\vspace*{1mm}

Physical quantities in QCD are independent of renormalization scheme (RS), 
but that exact invariance is spoiled by truncations of the perturbation 
series.  ``Optimization'' corresponds to making the perturbative approximant, 
at any given order, locally invariant under small RS changes.  A solution of 
the resulting optimization equations is presented.  It allows an efficient 
algorithm for finding the optimized result.  Example results for 
$R_{e^+e^-} = 3 \sum q_i^2 (1 + {\cal R})$ to fourth order (NNNLO) are given 
that show nice convergence, even down to arbitrarily low energies.  
The $Q=0$ ``freezing'' behaviour, ${\cal R} = 0.3 \pm 0.3$, found at third 
order is confirmed and made more precise; ${\cal R} =0.2 \pm 0.1$.  
Low-energy results in the $\msbar$ scheme, by contrast, show the 
typical pathologies of a non-convergent asymptotic series.  

\end{titlepage}

\setcounter{page}{1}

\section{Introduction}

      Renormalization, for physical quantities in massless QCD, amounts to 
eliminating the bare coupling constant in favour of a renormalized couplant. 
The precise definition of the renormalized couplant --- the renormalization 
scheme (RS) --- is in principle arbitrary, but at {\it finite} orders of 
perturbation theory the choice matters.  It is well known that one can 
take advantage of this situation by allowing the renormalization scale, $\mu$, 
to ``run'' with the experimental energy scale $Q$, but this familiar idea 
is vague and incomplete.  What matters is not $\mu$, in fact, but the ratio 
of $\mu$ to $\Lambda$, itself a RS-dependent parameter --- and, at higher 
orders, there are further sources of RS ambiguity.  ``Optimized perturbation 
theory'' (OPT) \cite{OPT} provides both a complete parametrization of RS 
dependence and the means to sensibly resolve these ambiguities by taking full 
advantage of Renormalization-Group (RG) invariance \cite{SP}.  

      The aim of this paper is twofold: ({\it i}) to present a mathematical 
solution to the ``optimization equations'' of Ref. \cite{OPT}, allowing the 
optimized result to be found efficiently; and ({\it ii}) to show numerical 
results for $R_{e^+e^-} \equiv \sigma(e^+e^- \to {\mbox{\rm hadrons}})/
\sigma(e^+e^- \to \mu^+ \mu^-)$, updating the results of 
Refs.~\cite{CKL,lowenlet,lowen} now that fourth-order calculations are 
available \cite{r3calc}.  

      The RS-dependence problem remains a controversial topic.  Our 
arguments have been set out in detail in Refs. \cite{OPT} and 
\cite{sense}--\cite{antiBLM}.  Here we give only a brief exposition of 
the key idea behind OPT, the ``principle of minimal sensitivity.''
\footnote{
  The importance and generality of this idea was emphasized in \cite{OPT} 
  but earlier authors had employed it in specific contexts, notably Caswell 
  and Killingbeck \cite{CK} in the anharmonic oscillator problem.  
}
This is the very general notion that, in any approximation method involving 
``extraneous'' parameters (parameters that one knows the exact result must 
be independent of), the sensible strategy is to find where the approximant 
is {\it minimally sensitive} to small variations of those parameters.  
The unknown exact result is globally invariant, while the approximant 
is not; where the approximant has the right qualitative behaviour --- local 
invariance (flatness as a function of the extraneous parameters) --- is where 
one can have most confidence in its quantitative value.  Many instructive 
examples testify to the basic soundness and power of this idea 
(see \cite{OPT},\cite{CK}--\cite{indcon}).

In the present context, RG invariance tells us that physical quantities 
should be independent of $\mu$ and all the other ``extraneous'' 
parameters involved in the RS choice.  That statement translates into an 
infinite set of equations (Eqs. (\ref{rga}) below) that any physical 
quantity ${\cal R}$ must satisfy \cite{OPT}.  Perturbative 
{\it approximations} to ${\cal R}$ do not satisfy these equations, but 
we can find an ``optimal RS'' in which they are satisfied locally.  
We explain in this paper how these optimization equations can 
be solved efficiently.  

Calculations in QCD perturbation theory, in particular for $\beta$ and 
$R_{e^+e^-}$, have progressed from leading order (LO) \cite{bcalc}, 
to next-to-leading order (NLO) \cite{ccalc,r1calc}, to 
next-to-next-to-leading (NNLO or N$^2$LO) \cite{c2calc, r2calc}, to 
now next-to-next-to-next-to-leading order (NNNLO or N$^3$LO) 
\cite{c3calc, r3calc}.  (We will use the terminology ``first order'' 
for leading order (LO) and ``$\kplus$ order'' for N$^k$LO.)  Tribute should 
be paid at this point to the heroic efforts involved in these enormously 
complex calculations.  The results allow us now to get a real sense of how QCD 
perturbation theory behaves, both in a fixed RS and with the optimization 
procedure.  

    Perturbation series, in any fixed RS, are expected to be factorially 
divergent.  However, it is possible that the optimized results converge, 
thanks to an ``induced convergence'' mechanism in which the optimized 
couplant shrinks from one order to the next \cite{optult} 
(see also \cite{Acol,rabin,indcon}).  Our numerical results here 
are quite consistent with that idea.

   The plan of this paper is as follows:  Section 2 reviews the mathematical 
consequences of RG invariance derived in Ref. \cite{OPT}.  Section 3 
discusses finite-order approximants and explains how to ``optimize'' the 
RS choice based on the principle of minimal sensitivity.  Section 4 solves 
the resulting optimization equations, in a general $\kplus$ order, 
and Section 5 outlines an algorithm to determine the result efficiently.    
Section 6 applies this algorithm to obtain illustrative results for 
$R_{e^+e^-}$ in second, third, and fourth orders and compares them with 
results in a fixed scheme, $\msbar(\mu=Q)$.  The optimized results 
exhibit steady convergence --- even down to zero energy, where the limit 
${\cal R} = 0.3 \pm 0.3$, found in third order~\cite{CKL,lowenlet,lowen}, 
is confirmed and made more precise; ${\cal R} =0.2 \pm 0.1$.  The 
fixed-scheme results show pathologies at low energies and we argue that 
these are a preview of what one can expect at higher orders at higher 
energies.  Concluding remarks are in Section 7.  Appendix A proves an 
identity mentioned in Section 4, and Appendix B discusses optimization in 
the fixed-point ($Q \to 0$) limit, following Ref. \cite{KSS}.

\section{RG invariance and its consequences}
\setcounter{equation}{0}

\subsection{RG equations}

     We begin by reviewing the formalism introduced in Ref.~\cite{OPT}.
First, we emphasize that OPT can only be applied to {\it physical} 
quantities (cross sections, decay rates, etc.).  QCD involves many other 
objects -- Green's functions, renormalized couplants, etc. -- that are 
{\it not} physically measurable and are {\it not} RS invariant; for our 
purposes these are merely intermediate steps in calculating physical 
quantities and need not be discussed.  In general, a perturbatively 
calculable physical quantity will have the form $A_1 {\cal R}+A_0$ with 
a leading-order coefficient $A_1$ and, sometimes, a zeroth-order term $A_0$.  
The coefficients $A_0$ and $A_1$, which carry dimensions of energy to the 
appropriate power, are RS invariant, so we may focus on dimensionless, 
normalized physical quantities ${\cal R}$ of the form:
\BE
{\cal R} = a^{\Pp}(1 + r_1 a + r_2 a^2 + \ldots),
\EE
where $a \equiv \alpha_s/\pi$ is the renormalized couplant.  
The power $\Pp$ is usually $1$ or $2$ or $3$, but need not be an integer.  
Generally a physical quantity ${\cal R}$ is not a single quantity but 
rather a function of several experimentally defined parameters.  We may always 
single out one such parameter with the dimensions of energy that we 
may call the ``experimental energy scale'' and denote by ``$Q$.''  (It is 
needed only to explain which quantities are, or are not, $Q$ dependent; the 
precise definition of $Q$ in any specific case is left to the reader.)  
Note that $Q$ must be defined in {\it experimental} terms and should not be 
confused with the renormalization scale $\mu$, which is defined only in 
terms of the technical details of the RS adopted.  

   In the specific case of $R_{e^+e^-}$ where, ignoring quark masses, 
\BE
R_{e^+e^-} = 3 \sum q_i^2 (1 + {\cal R}),
\EE
we have $\Pp=1$ and the $e^+e^-$ center-of-mass energy is a natural choice 
for $Q$. 

   The fundamental notion of RG invariance \cite{SP} means that a physical 
quantity is independent of the renormalization scheme (RS).  Expressed 
symbolically it states that
\begin{equation}
 0 = \frac{d{\cal R}}{d(RS)} =  
\left. \frac{\partial {\cal R}}{\partial (RS)} \right|_a  +
\frac{da}{d(RS)} \frac{\partial {\cal R}}{\partial a}, \label{rg} 
\end{equation}
where the total derivative is separated into two pieces corresponding, 
respectively, to RS dependence from the $r_i$ series coefficients, and from 
the couplant $a$ itself.  
A particular case of Eq. (\ref{rg}) is the familiar equation 
expressing the renormalization-scale independence of ${\cal R}$:
\begin{equation}
\left( \left. \mu \frac{\partial}{\partial \mu} \right|_a + 
\beta(a) \frac{\partial}{\partial a} \right) {\cal R} = 0,
\end{equation}
where 
\begin{equation}
\beta(a) \equiv \mu \frac{da}{d\mu} = - ba^2(1+ca + c_2a^2+c_3a^3+\ldots).
\end{equation}
The first two coefficients of the $\beta$ function are RS invariant 
\cite{tH} and, in QCD with $n_f$ massless flavours, are given by 
\cite{bcalc,ccalc}:
\begin{equation}
b = \frac{(33-2n_f)}{6}, \quad\quad c= \frac{153 - 19 n_f}{2(33-2n_f)}.
\end{equation}

     When integrated, the $\beta$-function equation can be written as:
\begin{equation}
\label{intbeta}
\int_0^{a} \frac{d a^{\prime}}{ \beta( a^{\prime})} + {\cal C} = 
\int_{\tilde{\Lambda}}^{\mu}
\frac{d \mu^{\prime}}{\mu^{\prime}} = \ln(\mu/{\tilde{\Lambda}}).
\end{equation}
where ${\cal C}$ is a suitably infinite constant and $\tilde{\Lambda}$ is a 
constant with dimensions of mass.  The particular definition of 
$\tilde{\Lambda}$ that we use corresponds to choosing \cite{OPT}
\begin{equation}
{\cal C} = \int_0^{\infty} \frac{d a^{\prime}}
{ b {a^{\prime}}^2(1 + c a^{\prime})}
\end{equation}
(where it is to be understood that the integrands on the left of 
(\ref{intbeta}) are to be combined before the bottom limit is taken).  This 
$\tilde{\Lambda}$ parameter is related to the traditional definition 
\cite{BBDM,PDG} by an RS-invariant, but $n_f$-dependent factor~\cite{OPT,Duke}:
\begin{equation}
\label{oldlam}
\ln (\Lambda/\tilde{\Lambda}) = (c/b) \ln (2c/b). 
\end{equation}

The $\tilde{\Lambda}$ parameter is scheme dependent, but the 
$\tilde{\Lambda}$-parameters of two schemes are related by the 
Celmaster-Gonsalves relation \cite{CG}.  
If two schemes are related, for the same value of $\mu$, by
\BE
\label{CG0}
a'=a(1+v_1 a+\ldots)
\EE
then
\BE
\label{CG0a}
\ln(\tilde{\Lambda}^{\prime}/\tilde{\Lambda}) = v_1/b.
\EE 
This relationship is {\it exact} and does not involve the $v_2, v_3, \ldots$ 
coefficients.  Thus, the $\tilde{\Lambda}$'s of different schemes can be 
related exactly by a 1-loop calculation.  Hence, the $\tilde{\Lambda}$ 
parameter of any convenient ``reference RS'' can be adopted, without 
prejudice, as the one free parameter of the theory, taking over the role 
of the ``bare coupling constant'' in the original Lagrangian.   

    From Eq. (\ref{intbeta}) it is clear that $a$ depends on RS only through 
the variables $\mu/\tilde{\Lambda}$ and $c_2, c_3, \ldots$, the 
scheme-dependent $\beta$-function coefficients.  The coefficients of 
${\cal R}$ can depend on RS only through these same variables --- the  
RG-invariance equation, (\ref{rg}), could not be satisfied otherwise.  
Therefore, these variables provide a complete RS parametrization, 
as far as physical quantities are concerned \cite{OPT}.  Thus, we may write:
\begin{equation}
 a=a({\rm RS})=a(\tau,c_2,c_3,\ldots), 
\end{equation}
where 
\begin{equation}
\tau \equiv b \ln(\mu/\tilde{\Lambda}).
\end{equation}
The $\tau$ variable is convenient and helps to emphasize the important 
point that $\mu$ itself is not meaningful because of the 
scheme ambiguity represented by Eq.~(\ref{CG0}); only the {\it ratio} of $\mu$ 
to $\tilde{\Lambda}$ matters.   

   The dependence of $a$ on the set of RS parameters $\tau$ and $ c_j$ 
is given by \cite{OPT}: 
\begin{equation} 
\frac{\partial a}{\partial \tau} = \beta(a)/b,
\end{equation}
\BE
\label{betaj}
\frac{\partial a}{\partial c_j} \equiv 
\beta_j(a) = -b \beta(a) \int_0^a dx \; \frac{x^{j+2}}{\beta(x)^2},
\EE
where the first equation is just the $\beta$-function equation in new 
notation, and the second follows by taking the partial derivative
of Eq. (\ref{intbeta}) with respect to $c_j$, holding $\tau$ and the 
other $c_i$ ($i \neq j$) constant. 

    The symbolic RG-invariance equation (\ref{rg}) can now be written out 
explicitly as the following set of equations~\cite{OPT}:
\BA
\label{rga}
 \frac{\partial {\cal R}}{\partial \tau} =
\left( \left. \frac{\partial}{\partial \tau} \right|_a + 
\frac{\beta(a)}{b} \frac{\partial}{\partial a} \right) {\cal R} \, = 0, 
 & \quad\quad ``j=1{\mbox{\rm''}} & \nonumber \\ 
 & & \\
\frac{\partial {\cal R}}{\partial c_j} =
\left( \left. \frac{\partial}{\partial c_j} \right|_a + 
\beta_j(a) \frac{\partial}{\partial a} \right) {\cal R} = 0 & \quad\quad 
j = 2,3,\ldots, & \nonumber
\EA

\subsection{The $\beta_j(a)$ and $B_j(a)$ functions}

    The $\beta_j$ functions, defined in Eq.~(\ref{betaj}), begin at order 
$a^{j+1}$ so it is convenient to define $B_j(a)$ functions whose series 
expansions begin $1+O(a)$:
\BE
B_j(a) \equiv \frac{(j-1)}{a^{j+1}} \beta_j(a).
\EE
For $j=1$ it is natural to define 
\BE
B_1(a) \equiv B(a) \equiv \frac{\beta(a)}{-b a^2} = 1+ c a+ c_2 a^2 + \ldots
 =  \sum_{i=0}^{\infty} c_i a^i,
\EE
with the convention that $c_0 \equiv 1$ and $c_1 \equiv c$.  
Equation (\ref{betaj}) can then be re-written as
\BE
\label{Bj}
B_j(a) = \frac{(j-1)}{a^{j-1}} B(a) I_j(a),
\EE
where 
\BE
\label{Ij}
I_j(a) \equiv \int_0^a dx \, \frac{x^{j-2}}{B(x)^2}.
\EE
(Note that this formula for $B_j(a)$ even holds for $j=1$ if the r.h.s. is 
interpreted as the limit $j \to 1$ from above.)

     The $B_j(a)$ functions have power-series expansions whose 
coefficients we write as $W^j_i$:
\BE
\label{Bjexp}
B_j(a) \equiv \sum_{i=0}^{\infty} W^j_i a^i,
\EE
with $W^j_0 \equiv 1$.  The other $W^j_i$ coefficients are fixed in terms 
of the $c_i$'s \cite{OPT}.  Differentiating Eq.~(\ref{betaj}), or 
equivalently by requiring commutation of the second 
derivatives, $\partial^2 a/\partial \tau \partial c_j = 
\partial^2 a/\partial c_j \partial \tau$, leads to \cite{OPT}
\BE
\label{diffeq}
\beta_j^{\prime}(a) \beta(a) - \beta^{\prime}(a) \beta_j(a) = - b a^{j+2},
\EE
where here the prime indicates differentiation with respect to $a$, regarding 
the coefficients $c_j$ as fixed.  
From this differential equation it is straightforward to show that the
$W^j_i$'s satisfy the relation
\BE
\label{Wcoeffs}
\sum_{m=0}^{i} (i+j-1-2m) c_m W^j_{i-m} = (j-1) \delta_{i0}
\EE
for $i=0,1,2,\ldots$ and $j=1,2,\ldots$.  In the special case $j=1$ one has 
$W^1_i \equiv c_i$ and the above equation reduces to 
\BE
\label{ccid}
\sum_{m=0}^{i} (i-2m) c_m c_{i-m} = 0,
\EE
which is true identically, since the left-hand side is
\BE
\sum_{m=0}^{i} (i-m) c_m c_{i-m} - \sum_{m=0}^{i} m c_m c_{i-m}
\EE
and the first sum, by changing the summation variable from 
$m$ to $n=i-m$, is seen to cancel the second.  

\subsection{The $\tilde{\rho}_n$ invariants ($\Pp=1$ case)}

\label{sec:invs}

The RG equations (\ref{rga}) determine how the coefficients $r_i$ of 
${\cal R}$ must depend on the RS variables $\{\tau,c_j\}$.  To show 
explicitly how this works we specialize to the $\Pp=1$ case, where 
\BE
{\cal R} = a(1+ r_1 a+ r_2 a^2 + \ldots)
\EE
and write out the lowest-order terms to obtain
\BE
\label{tauRG}
\left( a^2 \frac{\partial r_1}{\partial \tau} + 
a^3 \frac{\partial r_2}{\partial \tau} + \ldots \right)  
- a^2(1+ c a+ \ldots) (1+ 2 r_1 a+ \ldots)  =  0,
\EE
\BE
\label{c2RG}
\left( a^2 \frac{\partial r_1}{\partial c_2} + 
a^3 \frac{\partial r_2}{\partial c_2} + \ldots \right) 
+ a^3(1+ W^2_1 a+ \ldots) (1+ 2 r_1 a+ \ldots) =  0,
\EE
and so on.  (In fact, the coefficient $W^2_1$ is zero.)  
Equating powers of $a$, one sees that $r_1$ depends on $\tau$ only, while 
$r_2$ depends on $\tau$ and $c_2$ only, etc., with
\BE
\label{dr1dtau}
\frac{\partial r_1}{\partial \tau} = 1, 
\EE
\BE
\label{dr2}
\frac{\partial r_2}{\partial \tau} = 2 r_1 +  c,  \quad\quad 
\frac{\partial r_2}{\partial c_2} = - 1,  
\EE
etc..  Upon integration one will obtain $r_i$ as a function of 
$\tau,c_2,\ldots,c_i$ plus a constant of integration that is RS invariant.  
Thus, certain combinations of series coefficients and RS parameters are 
RS invariant \cite{OPT}.  The first two are
\BE
\label{rho1}
{\boldsymbol \rho}_1(Q) \equiv \tau - r_1,
\EE
\BE
\label{rho2}
\tilde{\rho}_2 \equiv  c_2 + r_2 - c r_1 - r_1^2.
\EE

     The first invariant, ${\boldsymbol \rho}_1(Q)$, is unique in being 
dependent on the experimental energy scale, $Q$.  A calculation of the 
coefficient $r_1$, in some arbitrary RS, yields a result of the form
\BE
r_1 = b \ln(\mu/Q) + r_{1,0},
\EE
whose $\mu$ dependence indeed conforms with Eq. (\ref{dr1dtau}).  For 
dimensional reasons, the $\mu$ and $Q$ dependences are tied together; 
the $r_1$ calculation does not ``know'' what boundary condition will later be 
applied to the $\beta$-function equation, so the parameter $\tilde{\Lambda}$ 
cannot explicitly appear.  Similarly, the higher coefficients $r_2, \ldots$ 
depend on $\ln(\mu/Q)$, but not on $\mu$ or $Q$ separately.  Hence, for the 
invariants $\tilde{\rho}_2, \tilde{\rho}_3, \ldots$ the cancellation 
of $\mu$ dependence also implies the cancellation of $Q$ dependence.  
However, ${\boldsymbol \rho}_1(Q)$ is different because its definition 
explicitly involves $\tau$, and we find
\BA
{\boldsymbol \rho}_1(Q) & = & b \ln(\mu/\tilde{\Lambda}) - 
\left( b \ln(\mu/Q) + r_{1,0} \right)
\nonumber \\
& = & b \ln (Q/\tilde{\Lambda}) - r_{1,0} 
\nonumber \\
& \equiv & 
b \ln (Q/\tilde{\Lambda}_{\cal R}),
\EA
where $\tilde{\Lambda}$ is the $\tilde{\Lambda}$-parameter in the RS in 
which the $r_1$ calculation was done, and $\tilde{\Lambda}_{\cal R}$ is a 
characteristic scale specific to the particular physical quantity ${\cal R}$.  
We can regard the last step as a Celmaster-Gonsalves relation, (\ref{CG0a}), 
that relates $\tilde{\Lambda}_{\cal R}$ back to the theory's one free 
parameter, the $\tilde{\Lambda}$ of some reference RS.  

     Some convention must be adopted to uniquely define the higher 
invariants  $\tilde{\rho}_j$ (for $j \ge 2$) because, of course, any sum 
of invariants is also an invariant.  For example, one might quite naturally 
add some multiple of $c^2$ to Eq. (\ref{rho2}).  Indeed, the tildes over the 
$\tilde{\rho}_j$'s are included to distinguish them from an earlier definition 
\cite{OPT}.  One convenient definition is as follows \cite{expform}.  
For any given physical quantity ${\cal R}$ one can always define a RS 
(known either as the ``fastest apparent convergence'' (FAC) or 
``effective charge'' \cite{Grunberg} scheme) such that all the series 
coefficients $r_i$ vanish in that scheme, so that 
${\cal R}=a_{\effch}(1+0+0+\ldots)$.  Since the $\beta$ functions of any 
two RS's are related by 
\BE
\label{betarel}
\beta'(a')  \equiv \mu \frac{da'}{d\mu} = \frac{da'}{da}  \mu \frac{da}{d\mu}
=   \frac{da'}{da} \beta(a),
\EE
we must have 
\BE
\beta_{\effch}({\cal R}) = \frac{d {\cal R}}{d a} \beta(a).
\EE
The $\tilde{\rho}_n$ invariants can be defined to coincide with the 
coefficients of the FAC-scheme $\beta$ function:
\BE
\beta_{\effch}({\cal R}) = - b {\cal R}^2 
\sum_{n=0}^\infty \, \tilde{\rho}_n {\cal R}^n = 
- b a^2 \frac{d {\cal R}}{d a} B(a)
\EE
where $\tilde{\rho}_0 \equiv 1$ and $\tilde{\rho}_1 \equiv c$ (not to be 
confused with the independent invariant 
$\boldsymbol{\rho}_1(Q) \equiv \tau - r_1$).  
Re-arranging this equation as
\BE
B(a) = \sum_{n=0}^\infty \, \tilde{\rho}_n a^n 
\left( \frac{{\cal R}}{a} \right)^{n+2} \frac{1}{ \frac{d {\cal R}}{d a} }
\EE
and equating powers of $a$ we obtain
\BE
\label{cjtorho}
c_j = \sum_{i=0}^{j} \, \tilde{\rho}_i \, \mathbb{C}_{j-i} \! 
\left[ \left( \frac{{\cal R}}{a} \right)^{i+2} 
\frac{1}{\frac{\partial {\cal R}}{\partial a}} \right],
\EE
where $\mathbb{C}_n [F(a)]$ means ``the coefficient of $a^n$ in the series 
expansion of $F(a)$.''  

  The first few invariants are listed below (for $\Pp=1$):
\BA
\tilde{\rho}_1 & = & c, \quad \quad {\mbox{\rm and}} \quad\quad 
\boldsymbol{\rho}_1(Q)= \tau-r_1 ,
\nonumber \\
\tilde{\rho}_2 & = & c_2+r_2-c r_1-r_1^2, 
\label{rhodefs}
\\
\tilde{\rho}_3 & = & c_3 + 2 r_3 -2 c_2 r_1 -6 r_2 r_1 + c r_1^2 + 4 r_1^3 .
\nonumber 
\EA
(Note that our earlier papers used a different convention, with 
$\rho_2^{\rm old} = \tilde{\rho}_2 - \quarter c^2$ and 
$\rho_3^{\rm old} = \half \tilde{\rho}_3$.)

\subsection{The $\tilde{\rho}_n$ invariants (general $\Pp$)}

    For general $\Pp$ the first few invariants are
\BA
\tilde{\rho}_1 & = & c, \quad \quad {\mbox{\rm and}} \quad\quad 
\boldsymbol{\rho}_1(Q)= \tau-\frac{r_1}{\Pp} ,
\nonumber \\
\tilde{\rho}_2 & = & c_2 + \frac{r_2}{\Pp} - \frac{c \, r_1}{\Pp} - 
\left( \frac{\Pp+1}{2 \Pp^2} \right) r_1^2, 
\label{rhodefsP}
\\
\tilde{\rho}_3 & = & c_3 + \frac{2 r_3}{\Pp} - \frac{2 c_2 r_1}{\Pp} 
-2 \left( \frac{\Pp+2}{\Pp^2} \right) r_2 r_1 + \frac{c \, r_1^2}{\Pp^2} 
+ \left( \frac{2(\Pp+1)(\Pp+2)}{3 \Pp^3} \right) r_1^3 .
\nonumber 
\EA

   The generalization of Eq.~(\ref{cjtorho}) to general $\Pp$ is given below 
in Eq.~(\ref{cjtorhoP}).  An explicit inverse formula giving the 
$\tilde{\rho}$'s in terms of the $c_j$ and $r_j$ coefficients can be found 
\cite{expform}:
\BE
\label{rhotocj}
\tilde{\rho}_n = \mathbb{C}_n \! 
\left[ B(a) \frac{1}{\Pp^2}\left( \frac{{\cal R}^{1/\Pp}}{a} \right)^{-(n+1)} 
\left( \frac{a}{{\cal R}} 
\frac{\partial {\cal R}}{\partial a} \right)^2 \right] .
\EE

\section{Finite orders and ``optimization''}
\setcounter{equation}{0}

\subsection{Finite-order approximants}

     So far our discussion has been formal and the results have been 
mathematical theorems.  Now we need to discuss finite-order approximants.  
At this point matters inevitably become controversial, because any 
approximation (unless it uses rigorously proven inequalities that bound the 
exact result) is necessarily a gamble; one is trying to guess at the exact 
result based on incomplete information.  The issue is how best to use all 
available information.  

   The first point to make is that {\it two} truncations are involved, for 
${\cal R}$ and for $\beta$.  (We need $\beta$ in order to relate $a$ to 
the $\tilde{\Lambda}$ parameter of some reference scheme.)  Thus, the 
$\kplus$ order, or (next-to)$^k$-leading order (N$^k$LO) approximant is 
naturally defined with both ${\cal R}$ and $\beta$ truncated after 
$k+1$ terms:
\BE
{\cal R}^{(k+1)} \equiv a^{\Pp} (1+ r_1 a + \ldots + r_k a^k),
\EE
where $a$ here is shorthand for $a^{(k+1)}$, the solution to the int-$\beta$ 
equation with $\beta$ replaced by $\beta^{(k+1)}$:
\BE
\beta^{(k+1)} \equiv -b a^2 (1+ c a + \ldots + c_k a^k).
\EE
It is straightforward to check that the order of the error term 
${\cal R}-{\cal R}^{(k+1)}$ is determined by whichever truncation, of 
${\cal R}$ or $\beta$, is the more severe, so it is natural 
to use the same number of terms in each \cite{OPT,sense}.  


In a fixed RS (with the RS choice also entailing a choice of $\mu$), the first 
step will be to find the value of $a$ in that RS by solving the integrated 
$\beta$-function equation, Eq.~(\ref{intbeta}).  That equation can be 
re-written (with $\tau \equiv b \ln(\mu/\tilde{\Lambda})$) in the form:  
\BE
\label{intbetaK}
\tau = K(a) = K^{(2)}(a) - \Delta(a),
\EE
where 
\BE
K^{(2)}(a) \equiv \int_a^\infty \frac{dx}{x^2(1+c x)} 
= \frac{1}{a} + c \ln \left| \frac{c a}{1+c a} \right| .
\EE
and
\BE
\Delta(a) \equiv  \int_0^a \! \frac{dx}{x^2} 
\left( \frac{1}{B(x)} - \frac{1}{1+c x} \right).
\EE
In $\kplus$ order $B(x)$ is replaced by $B^{(k+1)}(x) \equiv 
1+ c x + \ldots + c_k x^k$.  Hence $\Delta(a)$ would vanish in second order, 
so that $K^{(2)}(a)$ is indeed the second-order approximation to $K(a)$.  

   We shall refer to Eq. (\ref{intbetaK}) as the ``integrated 
$\beta$-function equation'' or ``int-$\beta$ equation.''  It should be 
solved {\it numerically} --- to an accuracy comfortably 
better than the expected error in the final result (see discussion in 
Section~\ref{sec:num}).  
To use an analytic approximation, such as a truncated expansion in inverse 
powers of $\ln(\mu^2/\Lambda^2)$ \cite{BBDM,PDG}, would introduce another 
uncontrolled approximation and create another source of ambiguity 
\cite{monsay,flconf}, namely dependence on how precisely the 
$\Lambda$-parameter is defined (e.g., the choice between $\tilde{\Lambda}$ 
and the more conventional $\Lambda$; see Eq.~(\ref{oldlam})). 
This is a wholly avoidable ambiguity and it is sensible to avoid it.

\subsection{Optimization in low orders ($\Pp=1$ case)}

\label{sec:optlow}

    While the exact ${\cal R}$ is RG-invariant, the finite-order approximants 
are not, since the truncations spoil the cancellations in the RG equations 
(\ref{rga}).   If ${\cal R}$ in those equations is replaced by 
${\cal R}^{(k+1)}$ then the r.h.s.~is not zero but is some remainder term 
$O(a^{\Pp+k+1})$.  As explained in the introduction, the idea of 
``optimized perturbation theory'' \cite{OPT} is to choose an ``optimal'' 
RS in which the approximant ${\cal R}^{(k+1)}$ is locally stationary with 
respect to RS variations; {\it i.e.}, the RS in which ${\cal R}^{(k+1)}$ 
satisfies the RG equations, (\ref{rga}), with no remainder:
\BA
\label{opteqs2}
\left( 
\left. \frac{\partial {\cal R}^{(k+1)}}{\partial \tau} \right|_a + 
\frac{\beta(a)}{b} \frac{\partial {\cal R}^{(k+1)}}{\partial a} 
\right)_{\rm opt. \, RS} = 0. 
& \quad\quad\quad &
{\mbox{\rm ``}}j=1{\mbox{\rm''}}
\nonumber \\
 & & \\
\left( 
\left. \frac{\partial {\cal R}^{(k+1)}}{\partial c_j} \right|_a + 
\beta_j(a) \frac{\partial {\cal R}^{(k+1)}}{\partial a} 
\right)_{\rm opt. \, RS} = 0. 
& \quad\quad\quad &
j=2,\ldots,k.
\nonumber
\EA

    We assume here that the QFT calculations of the ${\cal R}$ and 
$\beta$-function coefficients up to and including $r_k$ and $c_k$ have been 
done in some (calculationally convenient) RS.  From those results we can 
compute the values of the invariants $\boldsymbol{\rho}_1(Q)$ and 
$\tilde{\rho}_1 \equiv c$ and $\tilde{\rho}_2, \ldots, \tilde{\rho}_k$.  
Our optimized result will be expressed solely in terms of those invariants.  

    To see how this works let us consider the second-order 
(NLO) approximant. (For simplicity we set $\Pp=1$ for the
remainder of this subsection.)   
\BA
\label{R2}
{\cal R}^{(2)} & = & a (1 + r_1 a), \\
\beta^{(2)} & = & - b a^2 (1+ c a), 
\label{betatrun2}
\EA
where $a$ here is short for $a^{(2)}$, the solution to the int-$\beta$ 
equation (\ref{intbetaK}) with $\beta$ replaced by $\beta^{(2)}$:
\BE
\label{intbeta2}
\tau = K^{(2)}(a) = 
 \frac{1}{a} + c \ln \left| \frac{c a}{1+c a} \right|.
\EE
Since ${\cal R}^{(2)}$ depends on RS only through the variable $\tau$, only 
the ``$j=1$'' equation in Eq.~(\ref{opteqs2}) above is non-trivial.  Thus, 
the optimized ${\cal R}^{(2)}$ is determined by a single optimization 
equation:
\BE
\frac{\partial r_1}{\partial \tau} \, \bar{a}^2 - 
\bar{a}^2(1 + c \bar{a})(1 + 2 \bar{r}_1 \bar{a}) = 0.
\EE
(Overbars are used to indicate the value in the optimum RS.)  
As discussed in subsection~\ref{sec:invs}, the $a^2$ terms must cancel in 
any RS, which fixes $\frac{\partial r_1}{\partial \tau}=1$, leaving
\BE
1-(1+ c \bar{a})(1+ 2 \bar{r}_1 \bar{a})=0.
\EE 
This determines the optimized coefficient $\bar{r}_1$ in terms of the 
invariant $c$ and the optimized couplant $\bar{a}$:
\BE
\label{r1bar}
\bar{r}_1 = - \frac{1}{2} \frac{c}{1+ c \bar{a}}.
\EE
But $\bar{r}_1$ is related to $\bar{\tau}$ by the definition of the 
$\boldsymbol{\rho}_1(Q)$ invariant, Eq. (\ref{rho1}):
\BE
\boldsymbol{\rho}_1(Q) \equiv \bar{\tau} - \bar{r}_1.
\EE
Eliminating $\bar{r}_1$ between these last two equations and substituting into
the second-order int-$\beta$ equation, (\ref{intbeta2}), gives
\BE
\label{arho1eq}
 \frac{1}{\bar{a}} \left[  
1 + c \bar{a} \ln \left( \frac{c \bar{a}}{1+ c \bar{a}} \right) 
+ \frac{1}{2} \left( \frac{c \bar{a}}{1+ c \bar{a}} \right) \right] 
= \boldsymbol{\rho}_1(Q).
\EE
If the values of the invariants $c$ and $\boldsymbol{\rho}_1(Q)$ are known, 
as we assume, then we may numerically solve this last equation to obtain 
$\bar{a}$.   Substituting back in (\ref{r1bar}) we can find $\bar{r}_1$ and 
hence obtain the optimized approximant 
$\bar{{\cal R}}^{(2)}=\bar{a}(1+\bar{r}_1 \bar{a})$. 
 
   Note that the only approximations made here are the truncations of the 
${\cal R}$ and $\beta$ series in Eqs.~(\ref{R2},~\ref{betatrun2}),
which define the second-order approximant is some general RS.  We do not, for 
instance, approximate Eq. (\ref{r1bar}) as $\bar{r}_1 \approx -\half c$ 
(which corresponds to the PWMR \cite{PWMR} approximation, discussed later).  
Also, we will need to solve Eq.~(\ref{arho1eq}) numerically 

   We now turn to third order.  The third order approximant is defined by
\BA
{\cal R}^{(3)} & = & a(1 + r_1 a + r_2 a^2), \\
\beta^{(3)} & = & - b a^2 (1+ c a + c_2 a^2), 
\EA
where now $a$ is short for $a^{(3)}$, the solution to the int-$\beta$ 
equation with $\beta$ replaced by $\beta^{(3)}$.  Since ${\cal R}^{(3)}$ 
depends on RS through two parameters, $\tau$ and $c_2$, there are two 
optimization equations coming from Eq.~(\ref{opteqs2}).  These correspond to 
suitably truncated versions of Eqs.~(\ref{tauRG}, \ref{c2RG}).  Using 
Eqs.~(\ref{dr1dtau}, \ref{dr2}) they reduce to:
\BE
\label{eq1}
1+ \left( 2 \bar{r}_1 + c \right) \bar{a} 
- \bar{B}^{(3)}(\bar{a})
\left( 1+ 2 \bar{r}_1 \bar{a} + 3 \bar{r}_2 \bar{a}^2 \right) =0 ,
\EE
\BE
\label{eq2}
1 - \bar{B}_2^{(3)}(\bar{a})
\left( 1+ 2 \bar{r}_1 \bar{a} + 3 \bar{r}_2 \bar{a}^2 \right) =0 .
\EE
Here $\bar{B}^{(3)}(\bar{a}) \equiv 
\left( 1+ c \bar{a} + \bar{c}_2 \bar{a}^2 \right)$ is the $B(a)$ function 
truncated at third order in the optimum scheme.  The other function 
$\bar{B}_2^{(3)}(\bar{a})$ is not a polynomial; it is obtained from 
Eqs.~(\ref{Bj}, \ref{Ij}) with $B(x)$ replaced by $(1+ c x + \bar{c}_2 x^2)$. 
 
    These optimization equations, together with the definitions of the 
invariants $\tilde{\rho}_2$ and $\boldsymbol{\rho}_1(Q)$ and the 
int-$\beta$ equation, fully determine the optimized result, 
$\bar{{\cal R}}^{(3)}$ \cite{OPT,lowen}.  The new algorithm of Section 
\ref{sec:alg} provides a more efficient route to the result than the method 
in Ref.~\cite{lowen}.  
  
     Note that the ``optimum RS'' evolves from one order to the next; 
for instance $\bar{r}_1$ at third order is not the same as $\bar{r}_1$ at 
second order (so, strictly we should have distinguished $\bar{r}_1^{(2)}$ and 
$\bar{r}_1^{(3)}$ in the above).  

\subsection{The optimization equations} 

     We now write down the optimization equations at some general, $\kplus$, 
order \cite{OPT}.  To treat a general power $\Pp$, it is convenient to define
\BE
{\cal S} = \frac{1}{\Pp a^{\Pp-1}} \frac{\partial {\cal R}}{\partial a},
\EE
whose series expansion
\BE 
{\cal S} = 1+ s_1 a+ s_2 a^2 + \ldots
\EE
has coefficients 
\BE
s_m \equiv \left( \frac{m+\Pp}{\Pp} \right) r_m.
\EE
Using the $s_m$ coefficients absorbs all the $\Pp$ dependence in the 
analysis of the next section.  (However, $\Pp$ will re-appear later when we 
need to combine those results with the $\Pp$-dependent $\tilde{\rho}_n$ 
invariants.)

As we saw in subsection~\ref{sec:invs} (in the $\Pp=1$ case) all terms in 
the RG equations up to and including $O(a^{\Pp+k})$ must cancel automatically 
in any RS.  In the ``$j=1$'' optimization equation of Eq. (\ref{opteqs2}), 
the $\left. \frac{\partial {\cal R}^{(k+1)}}{\partial \tau}\right|_a$ 
term is a polynomial which must cancel the first $k$ terms of
$\frac{\beta(a)}{b} \frac{\partial {\cal R}^{(k+1)}}{\partial a}$.  
A similar observation applies to the other optimization equations of 
Eq.~(\ref{opteqs2}).  Hence, we may reduce the optimization conditions to 
\BE
\label{opteqs}
\bar{B}_j^{(k+1)}(\bar{a}) \bar{{\cal S}}^{(k+1)}(\bar{a}) - 
{\mathbb{T}}_{k-j} \left[ 
\bar{B}_j^{(k+1)}(\bar{a}) \bar{{\cal S}}^{(k+1)}(\bar{a}) \right] = 0,
\EE
for $j=1,2,\ldots,k$, where the notation 
$\mathbb{T}_n [F(a)]$ means ``truncate the series for 
$F(a)=F_0+F_1 a+\ldots$ immediately after the $a^n$ term''  
(i.e.,  $\mathbb{T}_n [F(a)] \equiv F_0 + F_1 a + \ldots + F_n a^n$.). 

     Note that the $B^{(k+1)}_j(a)$ functions are defined by 
Eqs. (\ref{Bj},~\ref{Ij}) with $B(a)$ replaced by the polynomial 
$B^{(k+1)}(a) \equiv 1+ c a + \ldots + c_k a^k$.

\section{Solution for the optimized $r_m$ coefficients}
\setcounter{equation}{0}

\subsection{Definition of the $H_i(a)$ functions}

     In this section it is implicit that all quantities are in the optimum 
RS at $\kplus$ order; overbars and ${\scriptstyle{(k+1)}}$ superscripts will 
be omitted.  
Also, we make the convention that 
\BE
r_0 \equiv s_0 \equiv c_0 \equiv 1, \quad {\mbox{\rm and}} \quad 
c_1 \equiv c.
\EE

      Next -- for reasons that will become clear in the next subsection 
-- we define some functions $H_1(a), \ldots, H_k(a)$ that are 
combinations of the $B_1(a), \ldots, B_k(a)$ functions:
\BE
\label{Hdef}
H_i(a) \equiv \sum_{j=0}^{k-i} c_j a^j \left( \frac{i-j-1}{i+j-1} \right) 
B_{i+j}(a) 
\quad\quad\quad i=(1),2,\ldots,k.
\EE
For $i=1$ this definition, as it stands, is ambiguous; it should be 
interpreted as 
\BE
\label{H1def}
H_1(a) = B_1(a) - \sum_{j=1}^{k-1} c_j a^j B_{j+1}(a),
\EE
corresponding to 
\BE
\label{ieq1}
\lim_{i \to 1} \left( \frac{i-j-1}{i+j-1} \right) =  \left\{ 
\begin{array}{ll}
1, \quad & j=0 \\
-1, \quad    & j \neq 0 
\end{array} \right.
\EE
It is also convenient and natural to define
\BE
H_0(a) \equiv 1 \quad \quad \mbox{{\rm and}} \quad\quad H_{k+1}(a) \equiv 0.
\EE

The $H$'s are defined as combinations of the $B$'s.  It turns out that 
there is a simple formula for the inverse relationship, giving the $B$'s 
as combinations of the $H$'s.

{\bf Lemma:}
\BE
\label{lemma}
B_j(a) = \sum_{q=0}^{k-j} W^j_q a^q H_{j+q}(a) \quad\quad\quad j=1,\ldots,k,
\EE
where the $W^j_i$ coefficients are those of the series expansion of 
$B_j(a)$, Eq. (\ref{Bjexp}).  (One might describe this result as follows: 
Take the power series for $B_j(a)$ and truncate it after the $a^{k-j}$ term.  
Now re-weight each term, replacing $a^q$ by $a^q H_{j+q}(a)$, and the result 
is the full series for $B_j(a)$.)

\vspace*{3mm}

{\bf Proof:} We first treat the cases with $j \neq 1$.  
Using the definition of the $H$'s, Eq.~(\ref{Hdef}), the r.h.s becomes 
\BE
\sum_{q=0}^{k-j} W^j_q a^q \sum_{p=0}^{k-j-q} c_p a^p 
\frac{(j+q-p-1)}{(j+q+p-1)} B_{j+q+p}(a).
\EE
Reorganizing the double sum by defining $n=q+p$ converts this expression to
\BE
\sum_{n=0}^{k-j} \frac{a^n B_{j+n}(a)}{(j+n-1)} 
\sum_{p=0}^n (n+j-1-2p) c_p W^j_{n-p}.
\EE
The inner sum reduces to $(j-1) \delta_{n0}$ by virtue of Eq.~(\ref{Wcoeffs}).
Thus, only the $n=0$ term of the outer sum survives, the $(j-1)$ factors 
cancel, and one is left with just $B_j(a)$, as claimed.  

    In the case $j=1$ the result to be proved, Eq.~(\ref{lemma}), becomes
\BE
B_1(a) = \sum_{q=0}^{k-1} c_q a^q H_{q+1}(a).
\EE
Using  Eq~(\ref{H1def}) for $H_1(a)$ and Eq~(\ref{Hdef}) for the other 
$H$'s, the r.h.s.  becomes
\BE
B_1(a) - \sum_{j=1}^{k-1} c_j a^j B_{j+1}(a) + \sum_{q=1}^{k-1} c_q a^q 
\sum_{j=0}^{k-q-1} c_j a^j \frac{(q-j)}{(q+j)} B_{q+j+1}(a).
\EE
Reorganizing the double sum by defining $n=q+j$ yields
\BE
\label{ieq1rhs}
B_1(a)  - \sum_{j=1}^{k-1} c_j a^j B_{j+1}(a) + 
\sum_{n=1}^{k-1} \frac{a^n}{n} B_{n+1}(a) \left( 
\sum_{q=1}^n (2q-n) c_q c_{n-q} \right)
\EE
The inner sum, in parentheses, after adding and subtracting a $q=0$ term 
becomes
\BE
\sum_{q=0}^n (2q-n) c_q c_{n-q} + n c_n = n c_n
\EE
since the full sum vanishes, as noted in Eq.~(\ref{ccid}).  Thus, the two 
series terms in (\ref{ieq1rhs}) cancel leaving just $B_1(a)$, as claimed.

\subsection{Formula for the optimized $s_m$ coefficients}

We are now ready to state the main new result; an exact, analytic expression 
for the optimized $s_m$ (and hence the $r_m$) coefficients, for 
$m=0,1,\ldots,k$, in terms of the (optimized values of) 
$a$ and the $\beta$-function coefficients $c_2, \ldots, c_k$:

{\bf Theorem}

The optimization equations (\ref{opteqs}) are satisfied by
\BE
\label{sform}
s_m = \frac{a^{-m}}{B_k(a)} \left( H_{k-m}(a) - H_{k-m+1}(a) \right), 
\quad\quad\quad m=0,1,\ldots,k.
\EE

\vspace*{3mm}

{\bf Proof:}
In the case $j=k$, where the $\mathbb{T}_0 [ \ldots ]$ term in (\ref{opteqs}) 
is just unity, the optimization equation reduces to 
\BE
{\cal S} = \frac{1}{B_k(a)}.
\EE
We first prove that this equation is satisfied.  Substituting Eq. 
(\ref{sform}) into the series for ${\cal S}$ gives
\BE
{\cal S} \equiv \sum_{m=0}^k  s_m a^m = 
\frac{1}{B_k(a)} \sum_{m=0}^k \left( H_{k-m}(a) - H_{k-m+1}(a) \right).
\EE
The $H$'s cancel in pairs leaving
\BE
{\cal S} = \frac{1}{B_k(a)} \left( H_0 - H_{k+1} \right)
=  \frac{1}{B_k(a)},
\EE
since we defined $H_0 \equiv 1$ and $H_{k+1} \equiv 0$ above.  

    Using this result and writing out the truncated-series term explicitly, 
the remaining optimization equations can be re-written as 
\BE
\label{opteqf}
\frac{B_j(a)}{B_k(a)} = \sum_{i=0}^{k-j}a^i \sum_{m=0}^i s_m W^j_{i-m}
\quad\quad\quad j=1,\ldots, k-1.
\EE
We now need to prove that these equations are satisfied by Eq. (\ref{sform}).
The r.h.s. becomes 
\BE
\sum_{i=0}^{k-j} a^i \sum_{m=0}^i W^j_{i-m} 
\frac{a^{-m}}{B_k(a)} \left( H_{k-m}(a) - H_{k-m+1}(a) \right).
\EE
Reorganizing the double summation, defining $q=i-m$ and thereby replacing 
$i$ with $m+q$ yields
\BE
\frac{1}{B_k(a)} \sum_{q=0}^{k-j} W^j_q a^q \sum_{m=0}^{k-j-q} 
\left( H_{k-m}(a) - H_{k-m+1}(a) \right).
\EE
The inner summation reduces to $H_{j+q}(a)$ since the $H$'s again cancel in 
pairs (and $H_{k+1} \equiv 0$).  Thus, the r.h.s of (\ref{opteqf}) reduces 
to 
\BE
\frac{1}{B_k(a)} \sum_{q=0}^{k-j} W^j_q  a^q  H_{j+q}(a) 
= \frac{B_j(a)}{B_k(a)},
\EE
where the last step uses the Lemma, Eq. (\ref{lemma}), and produces the 
l.h.s.~of (\ref{opteqf}), completing the proof.     

\subsection{An identity and the PWMR approximation}

\label{sec:idPWMR}

It is worth noting the following ``complete sum'' identity (proved in 
Appendix B) 
\BE
\sum_{j=0}^k \, c_j a^j \left( \frac{i-j-1}{i+j-1} \right) 
B_{i+j}(a) = 1
\quad\quad\quad i=(1),2,\ldots,k,
\EE
with the $i=1$ case interpreted using (\ref{ieq1}).  This identity reveals 
a remarkable property of the $H_i(a)$'s, which are defined as a 
``partial sum'' (over $j=0,\ldots,k-i$) of the same terms.  Hence, we can 
write
\BE
H_i(a) = 1- \sum_{j=k-i+1}^k c_j a^j \left( \frac{i-j-1}{i+j-1} \right) 
B_{i+j}(a)
\quad\quad\quad i=(1),2,\ldots,k \, ,
\EE
which, unlike the $H_i$ definition, involves $B_j$'s with $j$ greater than 
$k$.  Since the $B_j$'s all start $1+O(a)$ we see that the series for 
$(H_i(a)-1)$ begins only at order $a^{k-i+1}$:
\BE
H_i(a) - 1 = \frac{k-2i+2}{k} c_{k-i+1} a^{k-i+1} \left( 1+O(a) \right).
\EE
Substituting this result into Eq. (\ref{sform}) quickly leads to the result
\BE
s_m = \frac{k-2m}{k} c_m + O(a).
\EE
This result was first obtained --- in a quite different manner --- by 
Pennington, Wrigley, and Minaco and Roditi (PWMR) \cite{PWMR}.  The resulting 
PWMR approximation can be useful when $a \ll 1$.  (In fact, one also 
needs $a \ll a^*$ if a ``fixed point'' $a^*$ exists; see Appendix B.) \,
Inserting the above equation into the definitions of the $\tilde{\rho}_n$ 
invariants, one can find the PWMR-approximation $s_m$'s in terms 
of the invariants.  For $k=2$ one finds
\BE 
s_1 \approx 0, \quad s_2 \approx - \frac{3}{2} \tilde{\rho}_2,
\EE
while for $k=3$ one finds
\BE
s_1 \approx \frac{1}{3} c, \quad 
s_2 \approx -\frac{3}{8} \tilde{\rho}_2 - \frac{7}{96} c^2, \quad
s_3 \approx -2 \tilde{\rho}_3 - \frac{1}{2} \tilde{\rho}_2 c 
- \frac{1}{216} c^3. 
\EE
These results provide a useful starting point for the optimization algorithm 
described in the next section.

\section{Optimization algorithm}
\setcounter{equation}{0}

\label{sec:alg}

    The optimization problem at $\kplus$ order involves $2k+1$ variables, 
namely, $a$, $\tau, c_2, \ldots, c_k$, and $r_1, \ldots, r_k$.  These are 
connected by $2k+1$ equations, namely, the int-$\beta$ equation, the $k$ 
optimization equations, and the $k$ formulas for the invariants 
$\boldsymbol{\rho}_1(Q)$ and $\tilde{\rho_2}, \ldots, \tilde{\rho_k}$, 
whose numerical values we assume are given --- in the case of 
$\boldsymbol{\rho}_1(Q)$ the numerical value will depend on 
the value of $Q$ being considered.  We shall use $a, c_2, \ldots, c_k$ 
as the principal variables.  The solution to the optimization equations then 
explicitly determines the coefficients $r_1, \ldots, r_k$ in terms of these 
principal variables.  The int-$\beta$ equation explicitly fixes $\tau$ in 
terms of the principal variables.  The $\boldsymbol{\rho}_1(Q)= \tau-r_1$ 
equation can be used at the end to relate $a$ to $Q$, so the remaining task 
is to use the formulas for $\tilde{\rho_2}, \ldots, \tilde{\rho_k}$ to 
determine, self-consistently, by some convergent iterative procedure, the 
$c_2, \ldots, c_k$ variables.  One such algorithm is the following (recall 
$s_m \equiv \left( \frac{m+\Pp}{\Pp}\right) r_m$):

\vspace*{2mm}

\begin{quote}

{\bf (1)} Choose a numerical value for $a$.

{\bf (2)}  Make an initial guess for numerical values of 
$s_1, \ldots, s_k$ \\
(e.g. use the PWMR approximation). 

{\bf (3)} Find $c_j$ values from the invariants using 
Eq.~(\ref{cjtorho}) generalized to any $\Pp$:
\BE
\label{cjtorhoP}
c_j = \sum_{i=0}^{j} \, \tilde{\rho}_i \, \mathbb{C}_{j-i} \!
\left[ \Pp \left( \frac{{\cal R}^{1/\Pp}}{a} \right)^{i+1} 
\frac{1}{\frac{a}{{\cal R}} \frac{\partial {\cal R}}{\partial a}} \right],
\EE
where $\mathbb{C}_n [F(a)]$ means ``the coefficient of $a^n$ in the series 
expansion of $F(a)$.''

{\bf (4)}  Hence construct the $B(x)$ function as $\sum_{j=0}^{k} c_j x^j$ 
and then obtain $B(a)$ by substituting $x=a$ and the $B_j(a)$'s 
($j=2,\ldots, k$) by numerical integration of their definition:
\BE
\label{Bj4}
B_j(a) = \frac{(j-1)}{a^{j-1}} B(a) \int_0^a \! dx \frac{x^{j-2}}{B(x)^2}.
\EE

{\bf (5)}  Now find the $H_i(a)$'s from their definition:
\BE
\label{H5}
 H_i(a) = \sum_{j=0}^{k-i} \, c_j a^j 
\left( \frac{i-j-1}{i+j-1} \right) B_{i+j}(a) ,
\EE
and hence obtain new values for the $s_m$ coefficients from
\BE
\label{sm}
s_m = \frac{a^{-m}}{B_k(a)} \left( H_{k-m}(a) - H_{k-m+1}(a) \right) .
\EE

{\bf (6)}  Iterate from step {\bf 3} until the results converge to the 
desired precision.  

{\bf (7)}  Finally, use 
\BE
\boldsymbol{\rho}_1(Q) = \hat{K}^{(k+1)}(a)  - 
 \frac{s_1}{(\Pp+1)} 
\EE
(from the int-$\beta$ equation $\tau=\hat{K}^{(k+1)}(a)$, combined with 
$\boldsymbol{\rho}_1(Q) \equiv \tau-r_1/\Pp$) to find the value of $Q$ that 
corresponds to the chosen $a$ value. 

{\bf (8)}  
   One can then repeat the whole procedure with different initial $a$ values 
to cover the desired range of $Q$ values -- or to home in on one particular 
$Q$ value, if desired.  

\end{quote}
 
\vspace*{2mm}

Various details of this algorithm can be refined.  In particular one 
can try to avoid numerical problems with the large cancellation between the 
two $H$'s in (\ref{sm}), by constructing $H_i(a)-1$.  This algorithm appears 
to be quite robust and efficient provided that $B(a)$ is not too small; i.e., 
provided we are not close to a fixed point.  In that case one can find one 
of the iterations generating a negative $B(a)$ which makes the integrals 
in (\ref{Bj4}) blow up.  A simple cure for this problem is to is to use a 
modified algorithm that works with a fixed $B(a)$ rather than a fixed $a$.  
That is, one sets a fixed value of $B(a)=B_0$ at step {\bf 1} and then in 
step {\bf 4}, having constructed the new $B(x)$ function, one solves for a 
new $a$ from $B(a)=B_0$.  Also, near to the fixed point the PWMR approximation 
is not a good starting point and a better one is to use the fixed-point result 
instead (see Appendix B).

\section{Numerical Examples for $R_{e^+e^-}$}
\setcounter{equation}{0}

\label{sec:num}

\subsection{Procedure}

     We turn next to illustrative numerical results for a specific 
physical quantity, namely 
\BE
R_{e^+e^-} \equiv \frac{\sigma_{tot}(e^+e^- \rightarrow {\rm hadrons})}
{\sigma(e^+e^- \rightarrow \mu^+ \mu^-)}
\EE
at a centre-of-mass energy $Q$.  We shall neglect quark masses and consider 
$n_f$ flavours of quarks with electric charges $q_i$ ($i=1,\ldots,n_f$).  
If there were no QCD interactions then $R_{e^+e^-}$ would equal the 
parton-model result, $3 \sum q_i^2$.  Including perturbative QCD corrections 
we have 
\BE
R_{e^+e^-} = (3 \sum q_i^2)(1 + {\cal R}),
\EE
where ${\cal R}$ is a normalized physical quantity whose perturbation series 
has the form 
\BE
{\cal R} = a(1+ r_1 a + r_2 a^2 + \ldots).
\EE

    Previous discussions of this quantity to third order in OPT 
\cite{CKL,lowenlet,lowen} 
can now be extended to fourth order thanks to recent results from Baikov 
{\it et al} \cite{r3calc}.  Most importantly (see Example 7 below), we 
find that the infrared fixed-point (``freezing'') behaviour is confirmed 
and made more precise in fourth order.  There is, though, no need to update 
the phenomenological conclusions of Ref.~\cite{lowen} because the changes 
are well within the uncertainties discussed there.  Therefore, we shall 
ignore various phenomenological issues (the effect of quark masses, 
the matching of $\tilde{\Lambda}$'s at flavour thresholds \cite{marciano}, 
Poggio-Quinn-Weinberg smearing \cite{PQW}, experimental uncertainties, etc.) 
that were discussed in Ref.~\cite{lowen}.  

    Our focus here will be on the apparent convergence, or otherwise, of 
results from second to third to fourth order.  We shall compare the optimized 
results with those in a fixed scheme, the ``modified minimal subtraction'' 
($\msbar$) scheme with the renormalization scale $\mu$ chosen equal to the 
center-of-mass energy $Q$.  (Properly speaking, we should denote this 
scheme as $\msbar(\mu\!=\!Q)$, but for brevity we leave the $\mu\!=\!Q$ 
specification understood.)  We will essentially presume that the value of 
$\tilde{\Lambda}_\smmsbar$ is known from fitting other experimental data.   
However, to avoid committing to any specific value, we label our examples, 
not by $Q$, but by the ratio of $Q$ to $\tilde{\Lambda}_\smmsbar$.  At each 
order we proceed as if only the coefficients to that order had been 
calculated.  

   To obtain the $\msbar$ results, the first step is to evaluate the 
numerical value of the $\tau$ parameter of the $\msbar$ scheme:
\BE 
\label{tauMSbar}
\tau^\smmsbar = b \ln ( Q/\tilde{\Lambda}_\smmsbar ).  
\EE
Then, one must numerically solve the int-$\beta$ equation, (\ref{intbetaK}).  
At second order, where $B(a)$ is approximated by $1+c a$, this equation is 
\BE
\tau^\smmsbar = 
K^{(2)}(a) \equiv \frac{1}{a} + c \ln \left| \frac{c a}{1+c a} \right| .
\EE
With the resulting $a$, one then evaluates 
${\cal R}_{\smmsbar}^{(2)}=a(1+r_1^\smmsbar a)$.  At third order 
one must numerically solve
\BE
\label{intbetak3msbar}
\tau^\smmsbar = K^{(3)}_\smmsbar(a) \equiv K^{(2)}(a) - 
\Delta^{(3)}_\smmsbar(a),
\EE
where 
\BE
\Delta(a) \equiv  \int_0^a \! \frac{dx}{x^2} 
\left( \frac{1}{B(x)} - \frac{1}{1+c x} \right),
\EE
with, in this case, $B(x)$ approximated by $1+ c x + c_2 x^2$ with 
$c_2=c_2^\smmsbar$.  (A convenient approach is to first evaluate $\Delta$ at  
some initial value of $a$; then numerically solve Eq.~(\ref{intbetak3msbar}), 
with $\Delta$ treated as a constant, to obtain a new $a$; and then iterate.)  
With the resulting $a$ one then evaluates 
${\cal R}_{\smmsbar}^{(3)} = a(1+r_1^\smmsbar a+ r_2^\smmsbar a^2)$.  
At fourth order the procedure is the same, except that one now includes a 
$c_3^\smmsbar$ term in $B(x)$ and an $r_3^\smmsbar$ term in ${\cal R}$.   

     To obtain the optimized results one first needs to calculate the 
numerical values of the invariants.  At a given $Q/\tilde{\Lambda}_\smmsbar$ 
one can find $\tau^\smmsbar$ from Eq.~(\ref{tauMSbar}) and then obtain 
${\boldsymbol \rho}_1(Q)$ as $\tau^\smmsbar -r_1^{\smmsbar}$.  Numerical 
values for the $\tilde{\rho}_2$ and $\tilde{\rho}_3$ invariants (which are 
$Q$-independent) are similarly obtained by evaluating their definitions 
in Eq.~(\ref{rhodefs}) using the $\msbar$-scheme $r_i$ and $c_j$ coefficients. 
The optimized result to second order is obtained from Eqs.~(\ref{arho1eq}) 
and (\ref{r1bar}).  At higher orders one uses the algorithm described in 
the preceding section. 

    At each order one wants, not only a result for ${\cal R}$ but also 
an estimate for its likely error.  There is no rigorous way of doing this.  
However, it is reasonable to expect the ``apparent convergence'' of the 
series (i.e., the behaviour of the terms that have been calculated) is 
some sort of guide.  We shall adopt the common practice when dealing with 
asymptotic series of viewing the magnitude of the last calculated term, 
$\mid\! r_k a^{k+1} \!\mid$, as the error estimate.  We do this both for 
the $\msbar$ and the optimized results.  The change in the ${\cal R}$ results 
from one order to the next is another indicator; it seems quite consistent 
with our error estimate. 

    We will give two sets of examples; one set at moderately high energies, 
and the other at low energies.  For the first set of examples the appropriate 
number of flavours is $n_f=5$ ($u,d,s,c,b$ quarks), while for the second set 
it is $n_f=2$ ($u,d$ quarks only).

\subsection{High-energy examples}

For $n_f=5$ the $\beta$ function's leading, RS-invariant coefficients are
\BE
b=\frac{23}{6}, \quad\quad c= \frac{29}{23}.
\EE
In the $\msbar$ scheme its next two coefficients are \cite{c2calc,c3calc}
\BE
c_2^\smmsbar = \frac{9769}{6624} = 1.474789,
\EE
\BE
c_3^\smmsbar = -\frac{26017}{31104} + \frac{11027}{1242} \zeta(3) = 9.835916,
\EE
where $\zeta(s)$ is the Riemann zeta function.  
The $\msbar$ coefficients in ${\cal R}(e^+e^-)$ are 
\cite{r1calc,r2calc,r3calc}
\BE
r_1^\smmsbar=1.409230, \quad\quad r_2^\smmsbar=-12.80463, \quad\quad 
r_3^\smmsbar=-80.43373.
\EE
(The exact values, involving $\pi^2$, $\zeta(3)$, $\zeta(5)$ and $\zeta(7)$ 
\cite{r3calc} were used in our calculations.)  Inserting these values in 
Eq.~(\ref{rhodefs}) yields
\BE
\tilde{\rho}_2 = -15.0926, \quad\quad \tilde{\rho}_3 = -33.2216.
\EE

     Our first two examples update those in Table II of Ref.~\cite{lowen}.  
Results for $\kplus$ order ($k=1,2,3$) in both the $\msbar$ and optimized 
schemes are presented in the tables and figures below.  

   At these energies the perturbation series seems well behaved.  The 
$\msbar$ results are quite satisfactory but the optimized results offer 
greater precision, with smaller expected errors that tend to shrink more 
rapidly with increasing $k$.  It is noteworthy that while 
$a_\smmsbar$ slightly increases with $k$, the optimized couplant 
$\bar{a}$ shrinks, consistent with the ``induced convergence'' 
scenario of Ref.~\cite{optult}.

\newpage

\begin{center}

\textbf{Example 1:} 
$Q/\tilde{\Lambda}_{\scriptscriptstyle{\overline{\mathrm{MS}}}}=340$. 

\vspace*{4mm}

\begin{tabular}[b]{|l|l|l|l|}
\hline
Order & $a_{\scriptscriptstyle{\overline{\mathrm{MS}}}}$ & $\mathcal{R}_{%
\scriptscriptstyle{\overline{\mathrm{MS}}}}$ series & $\mathcal{R}_{%
\scriptscriptstyle{\overline{\mathrm{MS}}}}$ \\ \hline
$k=1$ & $0.0381237$ & $0.04(1+0.05)$ & $0.04017[205]$ \\ 
$k=2$ & $0.0382058$ & $0.04(1+0.05-0.02)$ & $0.03955[71]$ \\ 
$k=3$ & $0.0382161$ & $0.04(1+0.054-0.019-0.004)$ & $0.03939[17]$ \\ 
\hline\hline
Order & $\bar{a}$ & $\bar{\mathcal{R}}$ series & $\bar{\mathcal{R}}$ \\ 
\hline
$k=1$ & $0.0414570$ & $0.04(1-0.02) $ & $0.04043[103]$ \\ 
$k=2$ & $0.0394420$ & $0.04(1-0.01+0.01)$ & $0.03944[47]$ \\ 
$k=3$ & $0.0391507$ & $0.04(1+0.003+0.002+0.001)$ & $0.03941[4]$ \\ \hline
\end{tabular}

\vspace*{2mm}

\begin{quote}
Table 1.  Results for ${\cal R}$, the QCD corrections to $R_{e^+e^-}$, 
in $\kplus$ order (N$^k$LO) at an energy 
$Q/\tilde{\Lambda}_{\scriptscriptstyle{\overline{\mathrm{MS}}}}=340$.  
The upper and lower sub-tables list, respectively, the $\msbar$ and 
optimized results.  The columns give the couplant value, the rough 
form of the series, and the result for ${\cal R}$ with an error 
estimate corresponding to $\mid\! r_k a^{k+1} \!\mid$, 
the magnitude of the last term included in the perturbation series.
\end{quote}

\vspace*{2mm}

\begin{figure}[h]
\centering
\includegraphics[width=0.64 \textwidth]{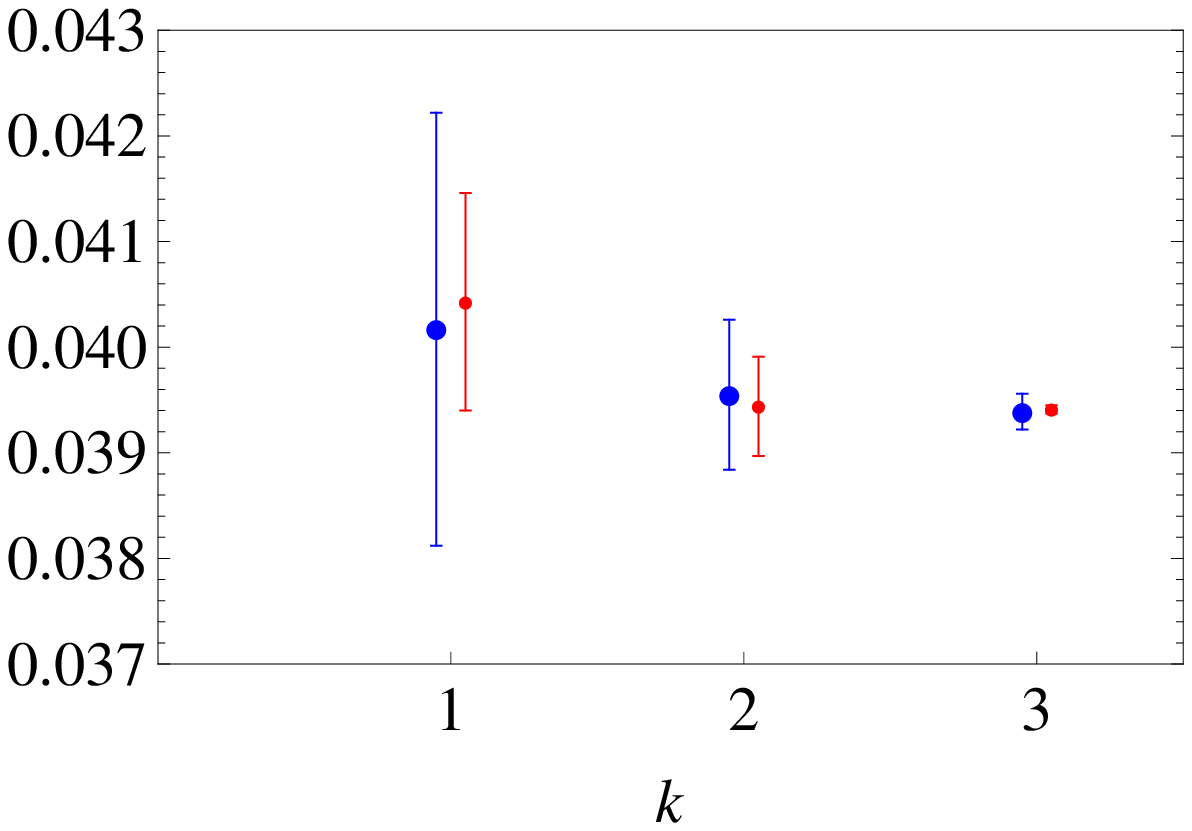}
\caption{Results for ${\cal R}$, the QCD corrections to $R_{e^+e^-}$,
in $\kplus$ order (N$^k$LO) at an energy 
$Q/\tilde{\Lambda}_{\scriptscriptstyle{\overline{\mathrm{MS}}}}=340$.  
The larger points displaced leftwards are in the $\msbar$ scheme, while 
the smaller points displaced rightwards are the optimized results.  
In both cases the error bars correspond to $\mid\! r_k a^{k+1} \!\mid$.}
\end{figure}

\end{center}


\begin{center}

\textbf{Example 2:} 
$Q/\tilde{\Lambda}_{\scriptscriptstyle{\overline{\mathrm{MS}}}}=68$. 

\vspace*{4mm}

\begin{tabular}[b]{|l|l|l|l|}
\hline
Order & $a_{\scriptscriptstyle{\overline{\mathrm{MS}}}}$ & $\mathcal{R}_{%
\scriptscriptstyle{\overline{\mathrm{MS}}}}$ series & $\mathcal{R}_{%
\scriptscriptstyle{\overline{\mathrm{MS}}}}$ \\ \hline
$k=1$ & $0.0507097$ & $0.05(1+0.07)$ & $0.05433[362]$ \\ 
$k=2$ & $0.0509032$ & $0.05(1+0.07-0.03)$ & $0.05287[169]$ \\ 
$k=3$ & $0.0509356$ & $0.05(1+0.07-0.03-0.01)$ & $0.05236[54]$ \\ 
\hline\hline
Order & $\bar{a}$ & $\bar{\mathcal{R}}$ series & $\bar{\mathcal{R}}$ \\ 
\hline
$k=1$ & $0.0568587$ & $0.06(1-0.03)$ & $0.05496[190]$ \\ 
$k=2$ & $0.0525541$ & $0.05(1-0.02+0.02)$ & $0.05256[112]$ \\ 
$k=3$ & $0.0520416$ & $0.05(1+0.002+0.003+0.002)$ & $0.05245[13]$ \\ \hline
\end{tabular}

\vspace*{2mm}

Table 2.  Results (as Table 1) for  
$Q/\tilde{\Lambda}_{\scriptscriptstyle{\overline{\mathrm{MS}}}}=68$.  

\vspace*{2mm}

\begin{figure}[h]
\centering
\includegraphics[width=0.64 \textwidth]{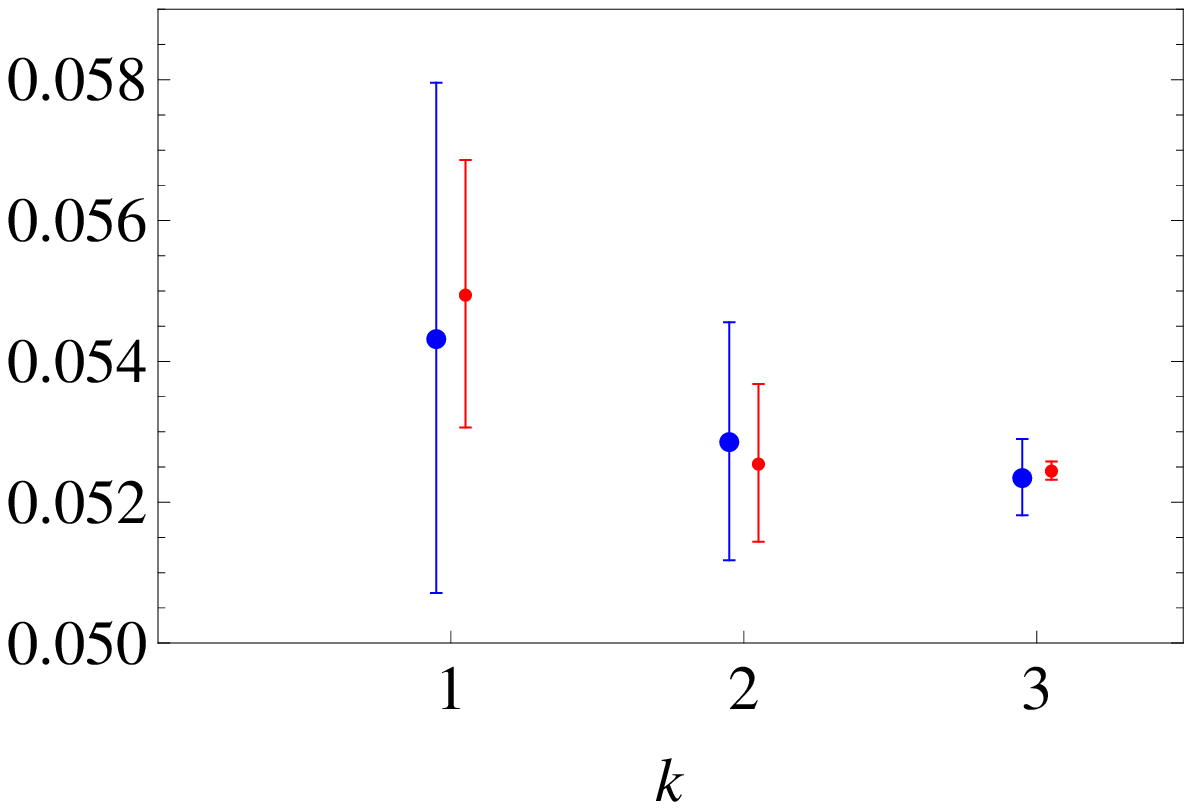}
\caption{As Fig.~1 but for  
$Q/\tilde{\Lambda}_{\scriptscriptstyle{\overline{\mathrm{MS}}}}=68$.}
\end{figure}

\end{center}

\newpage

\subsection{Low-energy examples}

Next we turn to lower-energy examples, where the differences between 
$\msbar$ and OPT are more dramatic.  With $n_f=2$ the $\beta$ function's 
leading coefficients are
\BE
b=\frac{29}{6}, \quad\quad c= \frac{115}{58},
\EE
and \cite{c2calc,c3calc}
\BE
c_2^\smmsbar = \frac{48241}{8352} = 5.77598,
\EE
\BE
c_3^\smmsbar = \frac{18799309}{902016} + \frac{68881}{12528} \zeta(3) 
= 27.45054.
\EE 
The ${\cal R}$ coefficients, in the $\msbar$ scheme, are 
\cite{r1calc,r2calc,r3calc}
\BE
r_1^\smmsbar=1.755117, \quad\quad r_2^\smmsbar=-9.14055, \quad\quad 
r_3^\smmsbar=-123.18799.
\EE
(Again, the exact values \cite{r3calc} were used in our calculations.)  
Inserting these values in Eq.~(\ref{rhodefs}) yields
\BE
\tilde{\rho}_2 = -9.92498, \quad\quad \tilde{\rho}_3 = -115.21021.
\EE

Tables 3--7 and Figures 3--7 give results at successively lower energies; 
$Q/\tilde{\Lambda}_\smmsbar=5, 2, 1.7, 1.5$ and $0$.  One sees in the 
$\msbar$ results the characteristic symptoms of an asymptotic series; 
after initially seeming to converge, the series starts to go bad, with the 
error estimate {\it increasing} with order.  In Example 3 the effect is 
just visible in the $k=3$ result, but in Examples 4 and 5 the effect becomes 
more dramatic.  In Example 6 there is no $k=3 \,\,\, \msbar$ result at all 
since for $Q/\tilde{\Lambda}_\smmsbar < 1.645$ there is no solution to the 
$k=3$ int-$\beta$ equation.  At still lower values of 
$Q/\tilde{\Lambda}_\smmsbar$, below $1.396$ and $1$ respectively, the
$k=2$ and $k=1$ $\msbar$ int-$\beta$ equations have no solution.  

By contrast, the optimized results show a monotonic decrease in the expected 
error at higher orders.  The $k=1$ results, in Examples 5 and 6 particularly, 
are very uncertain at low energies --- indeed, for 
$Q/\tilde{\Lambda}_\smmsbar < 1.438$ there is no solution to the $k=1$ 
optimal int-$\beta$ equation, Eq.~(\ref{arho1eq}).  However, for $k=2$ 
and $3$ the optimized results improve very significantly.  Indeed, for 
$k=2$ and $3$ there are optimized results down to zero energy, because 
the optimized $\beta$ function turns out to have a non-trivial fixed point 
(see Appendix B).  The new $k=3$ results provide solid confirmation of the 
earlier $k=2$ results \cite{CKL,lowenlet,lowen}.  As shown in the table for 
example 7, the $Q=0$ result for $\bar{{\cal R}}$ improves from $0.3\pm 0.3$ 
to $0.2\pm 0.1$.  

\newpage

\begin{center}

{\bf Example 3:}  $Q/\tilde{\Lambda}_\smmsbar=5$. 

\vspace*{4mm}

\begin{tabular}[b]{|l|l|l|l|}
\hline
Order & $a_\smmsbar$ & ${\cal R}_\smmsbar$ series &  ${\cal R}_\smmsbar$ \\
\hline
$k=1$ & $0.0862557$ & $0.09(1+0.15)$ & $0.099[13]$ \\
$k=2$ & $0.0902494$ & $0.09(1+0.16-0.07)$ & $0.098[7]$ \\
$k=3$ & $0.0911287$ & $0.09(1+0.16-0.08-0.09)$ & $0.090[8]$ \\
\hline \hline
Order & $\bar{a}$ & $\bar{{\cal R}}$ series & $\bar{{\cal R}}$ \\
\hline
$k=1$ & $0.117285$ & $0.12(1-0.09)$ & $0.106[11]$ \\
$k=2$ & $0.0952429$ & $0.10(1-0.05+0.05)$ & $0.095[5]$ \\
$k=3$ & $0.0899359$ & $0.09(1-0.01-0.02+0.04)$ & $0.091[4]$ \\
\hline 
\end{tabular}

\vspace*{2mm}

Table 3.  Results for $Q/\tilde{\Lambda}_\smmsbar=5$. 

\vspace*{2mm}

\begin{figure}[h]
\centering
\includegraphics[width=0.64 \textwidth]{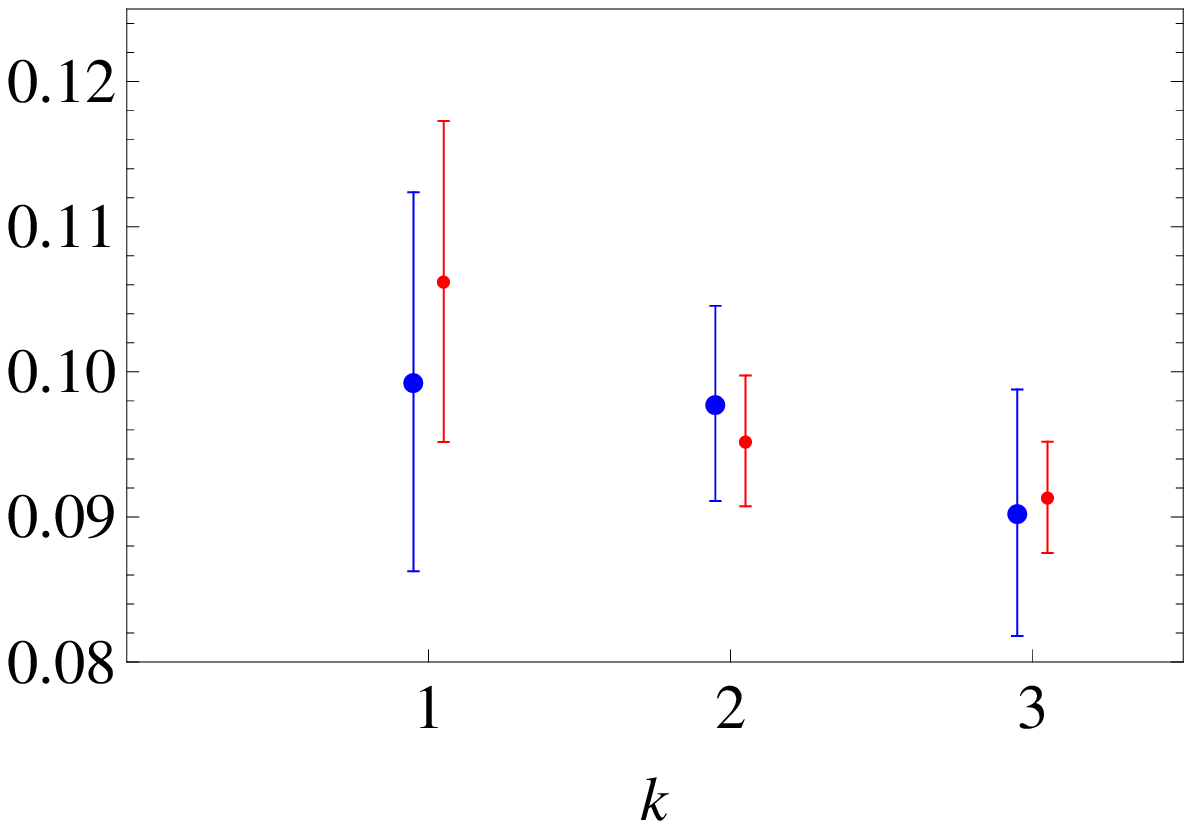}
\caption{As Fig. 1 but for $Q/\tilde{\Lambda}_\smmsbar=5$.}
\end{figure}

\end{center}

\newpage

\vspace*{3mm}

\begin{center}

{\bf Example 4:}  $Q/\tilde{\Lambda}_\smmsbar=2$. 

\vspace*{4mm}

\begin{tabular}[b]{|l|l|l|l|}
\hline
Order & $a_\smmsbar$ & ${\cal R}_\smmsbar$ series &  ${\cal R}_\smmsbar$ \\
\hline
$k=1$ & $0.1626471$ & $0.16(1+0.29)$ & $0.209[46]$ \\
$k=2$ & $0.1963533$ & $0.20(1+0.34-0.35)$ & $0.195[69]$ \\
$k=3$ & $0.2193679$ & $0.22(1+0.39-0.44-1.30)$ & $-0.08 \pm 0.29$ \\
\hline \hline
Order & $\bar{a}$ & $\bar{{\cal R}}$ series & $\bar{{\cal R}}$ \\
\hline
$k=1$ & $0.3648099$ & $0.36(1-0.21)$ & $0.288[77]$ \\
$k=2$ & $0.1725913$ & $0.17(1-0.17+0.18)$ & $0.173[31]$ \\
$k=3$ & $0.1421756$ & $0.14(1-0.08-0.09+0.17)$ & $0.143[25]$ \\
\hline 
\end{tabular}

\vspace*{2mm}

Table 4.  Results for $Q/\tilde{\Lambda}_\smmsbar=2$. 

\vspace*{2mm}

\begin{figure}[h]
\centering
\includegraphics[width=0.64 \textwidth]{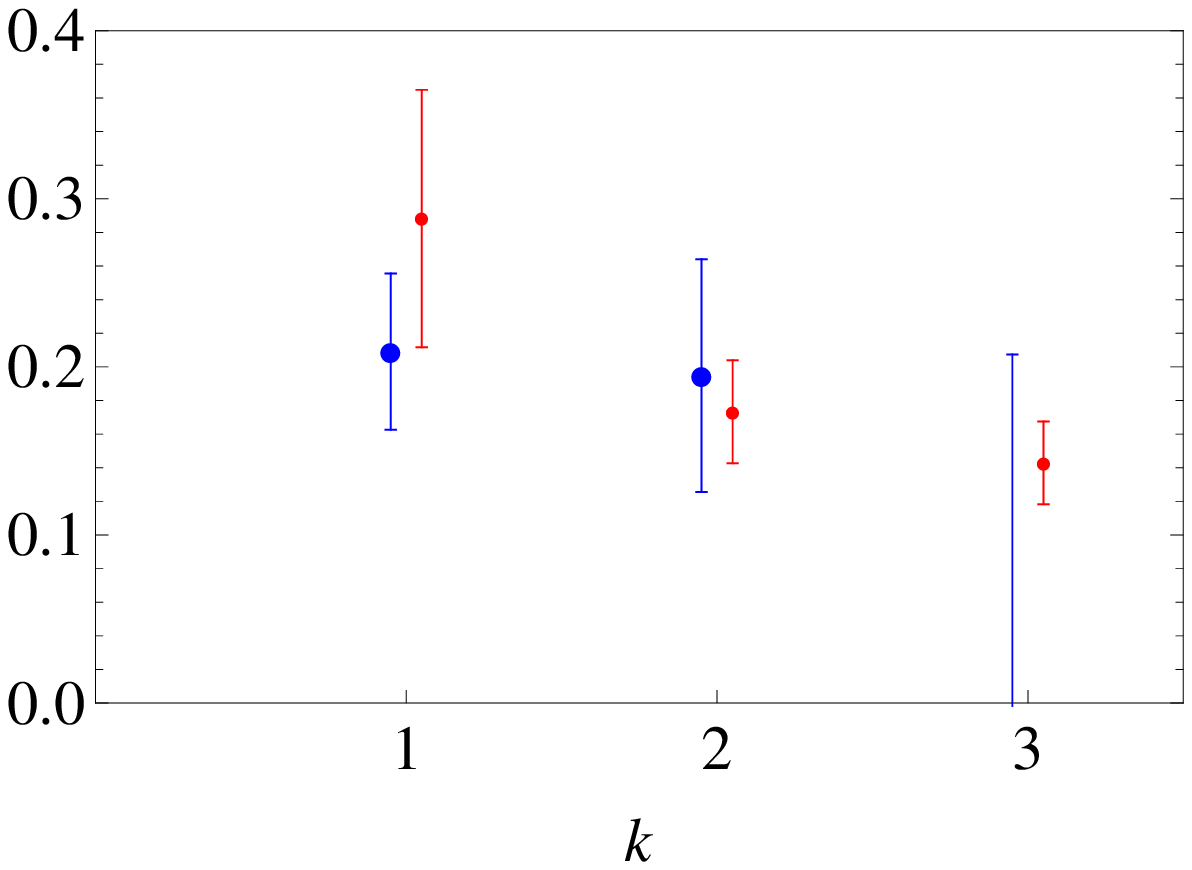}
\caption{As Fig. 1 but for $Q/\tilde{\Lambda}_\smmsbar=2$.  (The 
$k=3 \,\,\, \msbar$ result is slightly negative; only its error bar is 
visible.)}
\end{figure}

\vspace*{2mm}

\end{center}  

\newpage

\vspace*{3mm}

\begin{center}

{\bf Example 5:}  $Q/\tilde{\Lambda}_\smmsbar=1.7$. 

\vspace*{4mm}

\begin{tabular}[b]{|l|l|l|l|}
\hline
Order & $a_\smmsbar$ & ${\cal R}_\smmsbar$ series &  ${\cal R}_\smmsbar$ \\
\hline
$k=1$ & $0.1966624$ & $0.20(1+0.35)$ & $0.265[68]$ \\
$k=2$ & $0.2691684$ & $0.27(1+0.47-0.66)$ & $0.218[178]$ \\
$k=3$ & $0.4153849$ & $0.42(1+0.73-1.58-8.83)$ & $-3.60 \pm 3.67$ \\
\hline \hline
Order & $\bar{a}$ & $\bar{{\cal R}}$ series & $\bar{{\cal R}}$ \\
\hline
$k=1$ & $0.6669931$ & $0.67(1-0.28)$ & $0.477[190]$ \\
$k=2$ & $0.1970393$ & $0.20(1-0.24+0.25)$ & $0.199[49]$ \\
$k=3$ & $0.1530735$ & $0.15(1-0.11-0.11+0.22)$ & $0.153[34]$ \\
\hline 
\end{tabular}

\vspace*{2mm}

Table 5.  Results for $Q/\tilde{\Lambda}_\smmsbar=1.7$.

\vspace*{2mm}

\begin{figure}[h]
\centering
\includegraphics[width=0.64 \textwidth]{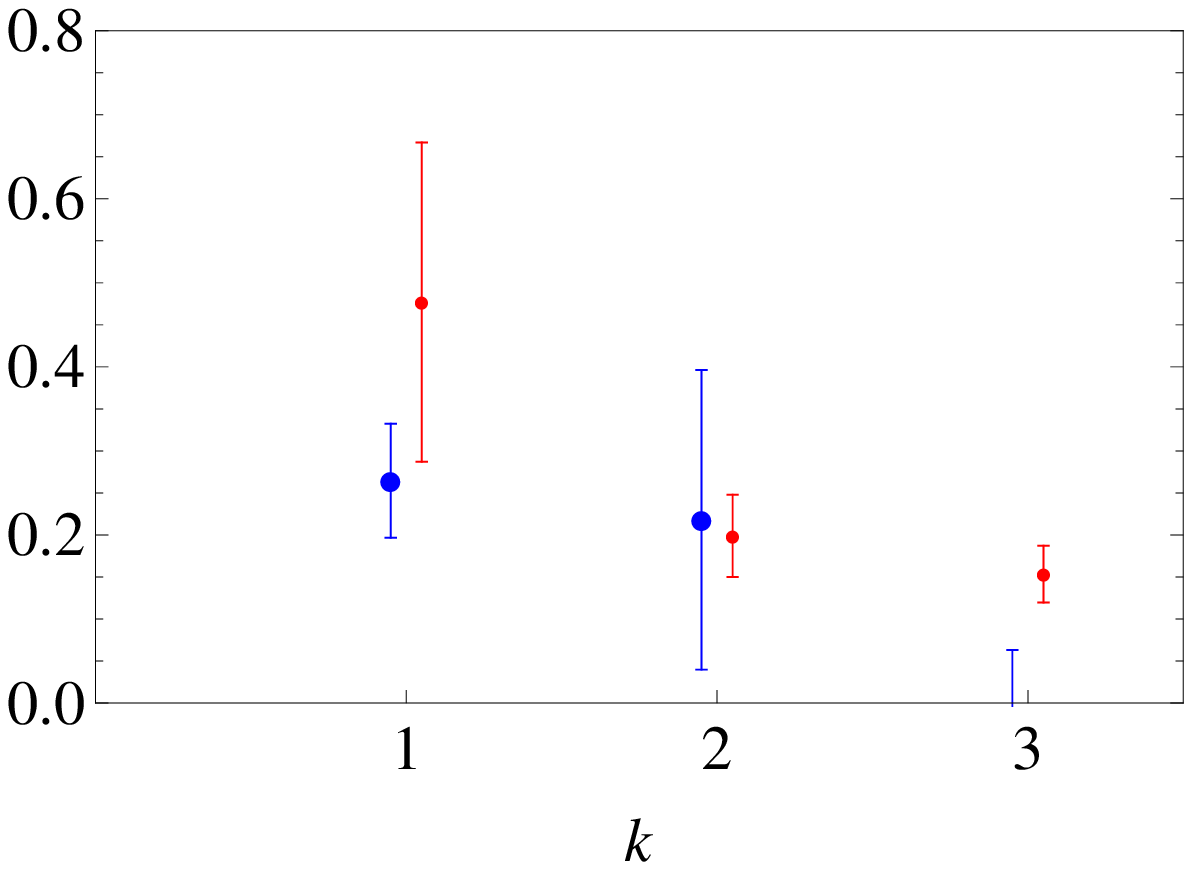}
\caption{As Fig. 1 but for $Q/\tilde{\Lambda}_\smmsbar=1.7$.  (Only 
the tip of the huge error bar for the $k=3 \,\,\, \msbar$ result is visible.)}
\end{figure}

\vspace*{2mm}

\end{center}  

\newpage

\vspace*{3mm}

\begin{center}

{\bf Example 6:}  $Q/\tilde{\Lambda}_\smmsbar=1.5$. 

\vspace*{4mm}

\begin{tabular}[b]{|l|l|l|l|}
\hline
Order & $a_\smmsbar$ & ${\cal R}_\smmsbar$ series &  ${\cal R}_\smmsbar$ \\
\hline
$k=1$ & $0.236877$ & $0.24(1+0.42)$ & $0.335[98]$ \\
$k=2$ & $0.431322$ & $0.43(1+0.76-1.70)$ & $0.02 \pm 0.73$ \\
$k=3$ & no solution &  &  \\
\hline \hline
Order & $\bar{a}$ & $\bar{{\cal R}}$ series & $\bar{{\cal R}}$ \\
\hline
$k=1$ & $2.4690661$ & $2.5(1-0.42)$ & $1.4\pm1.0$ \\
$k=2$ & $0.2173977$ & $0.22(1-0.31+0.33)$ & $0.221[71]$ \\
$k=3$ & $0.1605183$ & $0.16(1-0.13-0.13+0.26)$ & $0.161[42]$ \\
\hline 
\end{tabular}

\vspace*{2mm}

Table 6.  Results for $Q/\tilde{\Lambda}_\smmsbar=1.5$. 

\vspace*{2mm}

\begin{figure}[h]
\centering
\includegraphics[width=0.64 \textwidth]{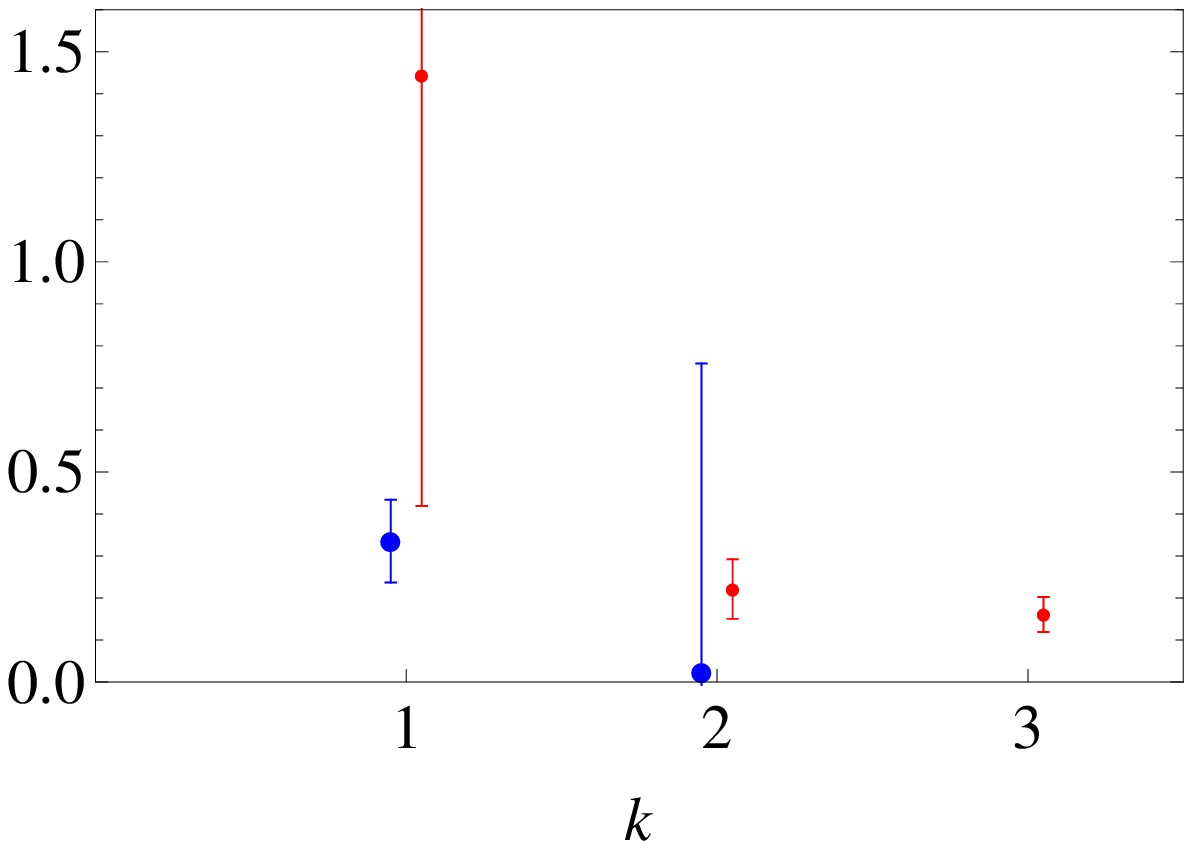}
\caption{As Fig. 1 but for $Q/\tilde{\Lambda}_\smmsbar=1.5$.  
(There is no $k=3 \,\,\, \msbar$ result in this case.)}
\end{figure}

\end{center}  

\newpage

\vspace*{3mm}

\begin{center}

{\bf Example 7:}  $Q/\tilde{\Lambda}_\smmsbar=0$ \quad (fixed point).

\vspace*{4mm}

\begin{tabular}[b]{|l|l|l|l|}
\hline
Order & $\bar{a}$ & $\bar{{\cal R}}$ series & $\bar{{\cal R}}$ \\
\hline
$k=1$ & no solution &  &  \\
$k=2$ & $0.2635259$ & $0.26(1-0.76+1.01)$ & $0.330[267]$ \\
$k=3$ & $0.1800794$ & $0.18(1-0.25-0.16+0.44)$ & $0.185[79]$ \\
\hline 
\end{tabular}

\vspace*{2mm}

\begin{quote}
Table 7.  Results for the infrared fixed-point limit, 
$Q/\tilde{\Lambda}_\smmsbar=0$.  There are no $\msbar$ results in this 
case.
\end{quote}

\vspace*{2mm}

\begin{figure}[h]
\centering
\includegraphics[width=0.64 \textwidth]{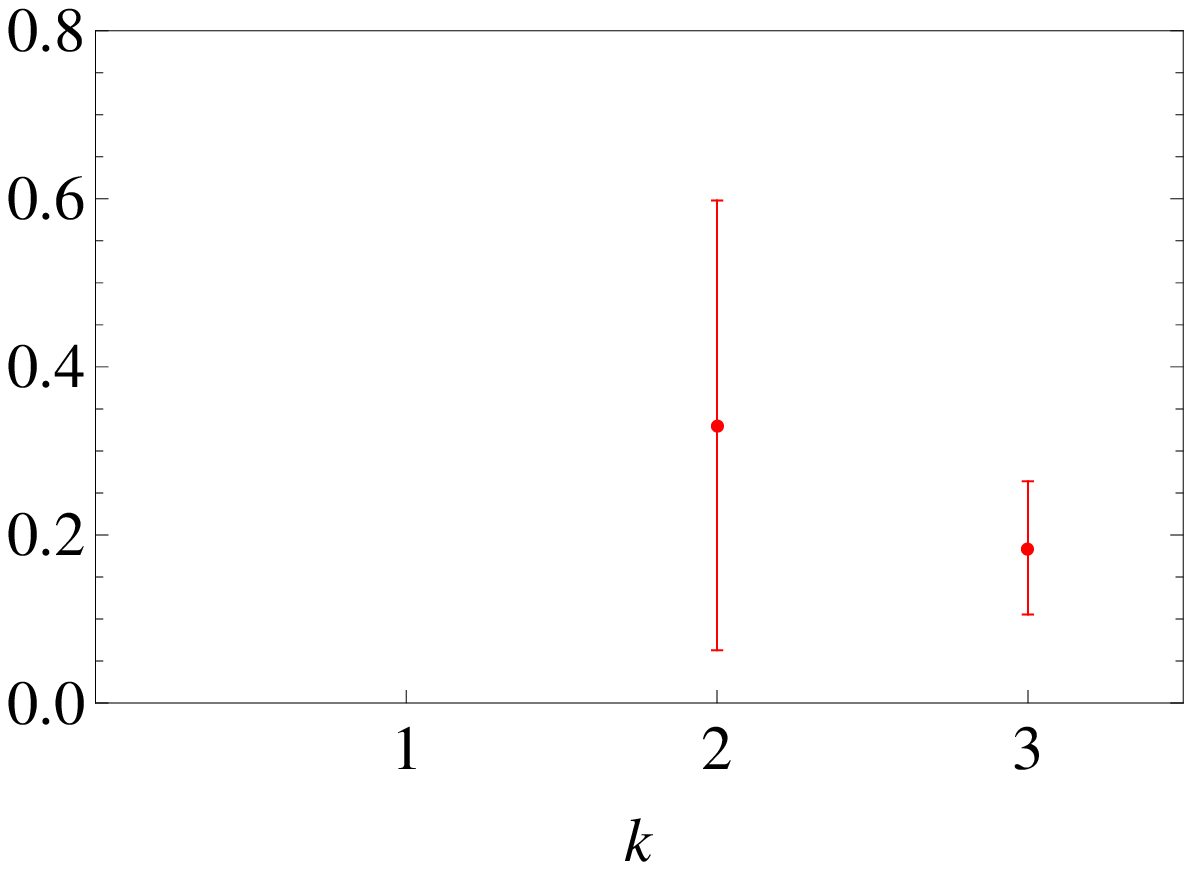}
\caption{As Fig. 1 but for the infrared fixed-point limit, 
$Q/\tilde{\Lambda}_\smmsbar=0$.  There are no $\msbar$ results in this 
case, and no $k=1$ optimized result.}
\end{figure}

\end{center}  

\newpage

\section{Concluding Remarks}
\setcounter{equation}{0}

   We first summarize the lessons of the numerical examples, which compared 
the $\msbar(\mu=Q)$ and optimized results.  At moderately high energies, 
the differences are well within the error estimates; the main 
advantage of optimization here is to achieve greater precision.  At low 
energies, however, the optimized results show steady convergence, while 
the $\msbar$ results begin to show the typical pathologies of a 
non-convergent asymptotic series.  We can expect these pathologies to 
eventually show up in the $\msbar$ results at higher energies when the series 
is taken to high enough order.  In this sense the low-energy examples are a 
``preview'' of the divergent-series problems to be expected in $\msbar$ or 
any fixed RS.   

     Whether or not the optimized results ultimately converge is a matter 
of conjecture at present.  However, the lessons of toy models 
\cite{optult,Acol} and from the linear $\delta$-expansion for the anharmonic 
oscillator \cite{indcon}, together with the present results, suggest that 
it is a very real possibility.  We should say at once that we would {\it not} 
expect the optimized results to converge to the exact, physical answer.  
We certainly expect there to be nonperturbative contributions (higher-twist 
terms that involve powers of $e^{-1/(b a)}$ and hence are invisible to 
perturbation theory).  A physically defined quantity ${\cal R}$ in QCD is, 
in general, a sum of perturbative and nonperturbative terms, 
${\cal R}_{\rm pert} + {\cal R}_{\rm nonpert}$.  The issue, though, is 
whether this decomposition can be made unique and physically meaningful 
or whether it is inherently ambiguous and dependent on RS.  It seems 
reasonable that ${\cal R}_{\rm nonpert}$ should involve only infrared 
physics and so should be calculable, in principle, without any essential 
need for renormalization.  If that is so, there would have to be a 
version of perturbation theory that converges to a RS-invariant result, 
${\cal R}_{\rm pert}$.  Optimization, where at each order we demand that 
the result be ``as RS invariant as possible,'' is the natural candidate to 
be that version of perturbation theory --- without any need to invoke 
extra tricks (such as Pad\'e approximants, Borel summation, etc.).

     The good convergence of the optimized results in the $e^+e^-$ case 
remains true even down to zero energy, where the third-order finding 
\cite{CKL,lowenlet,lowen} of a limit ${\cal R} = 0.3 \pm 0.3$ is confirmed 
and made more 
precise; ${\cal R} =0.2 \pm 0.1$.  This result is very important because there 
are many indications from phenomenology (see the mini-review in 
Ref.~\cite{lowen}) that the QCD couplant does freeze at low energies.  
Usually freezing is something that theorists put in by hand, but here it is 
an outcome, a {\it prediction}.  There is nothing in the optimization approach 
that forces freezing to occur; the fact that it does for $R_{e^+e^-}$ is due 
to the $\tilde{\rho}_2, \tilde{\rho}_3$ values resulting from the 
Feynman-diagram calculations.  

     The methods described here could be applied to many other perturbative 
physical quantities.  In particular, fourth-order results are now 
available for $\tau$-lepton and $Z^0$ decay widths and for some 
deep-inelastic-scattering sum rules \cite{r3calc,Bjsum}.  Previous studies 
of OPT applied to these quantities \cite{CKL,Matt,ChyKat} could now be 
extended to fourth order.  

    We have not discussed here physical quantities that explicitly 
involve parton distribution functions or fragmentation functions.  
Such quantities are plagued by another kind of ambiguity; 
factorization-scheme dependence --- an ambiguity similar to, and 
entangled with, RS dependence.  The ``principle of minimal sensitivity'' 
can be applied here, too.  Unfortunately, the original analysis 
\cite{Politzer} was stymied by an algebraic error, corrected later 
in Ref.~\cite{stpol} (see also \cite{NNY}).  These papers use the language of 
structure-function moments --- which, while natural theoretically, is perhaps 
not very convenient phenomenologically.  A purely numerical approach to 
``optimization'' \cite{aurenche,chyla} is certainly feasible, but laborious.  
It would be valuable to somehow reformulate Ref.~\cite{stpol}'s results 
in a way that would combine easily with practical methods for using and 
empirically determining parton distribution functions.  
We suspect, based on the important results of \cite{aurenche,chyla}, that 
several important QCD cross sections are currently underestimated, and that 
``optimization'' could significantly reduce the theoretical uncertainties of 
many others.

\begin{center}
{\bf Acknowledgments}
\end{center}
\begin{quote}
I am grateful to T. J. Sarkar for discussions that were helpful in refining 
the solution for the optimized coefficients.  I also thank Tal Einav for 
detailed comments on an early version of the manuscript.  
\end{quote}

\section*{Appendix A:  Proof of the ``complete sum'' identity}
\renewcommand{\theequation}{A.\arabic{equation}}
\setcounter{equation}{0}

     This appendix provides a proof of the identity, mentioned in 
subsection~\ref{sec:idPWMR}, for the ``complete sum:''
\BE
\label{completesum}
\sum_{j=0}^k \, c_j a^j \left( \frac{i-j-1}{i+j-1} \right) 
B_{i+j}(a) = 1
\quad\quad\quad i=(1),2,\ldots,k.
\EE

     From the definition of the $\beta_j(a)$ and $B_j(a)$ functions we have 
Eqs. (\ref{Bj}) and (\ref{Ij}).  Substituting 
into the l.h.s. of (\ref{completesum}) yields
\BE
\label{completesum2}
\frac{B(a)}{a^{i-1}} \sum_{j=0}^k \, c_j (i-j-1) 
\int_0^a \! dx \frac{x^{i+j-2}}{B(x)^2}.
\EE
Pulling the integration outside the sum and re-grouping leads to
\BE
\frac{B(a)}{a^{i-1}} \int_0^a \! dx \frac{x^{i-2}}{B(x)^2} \left(
(i-1) \sum_{j=0}^{k} c_j x^j - \sum_{j=0}^{k} j c_j x^j \right).
\EE
The integrand can now be recognized as 
\BE
\frac{x^{i-2}}{B(x)^2} \left( (i-1) B(x) - x \frac{dB(x)}{dx} \right)
= \frac{d}{dx} \left( \frac{x^{i-1}}{B(x)} \right),
\EE
so the integral can be done immediately, giving
\BE
\frac{B(a)}{a^{i-1}} \left[ \frac{x^{i-1}}{B(x)} \right]_{x=0}^{x=a}.
\EE
For $i>1$ the lower endpoint ($x=0$) makes no contribution, so the result is 
unity, as claimed.  

      The case $i=1$ requires special treatment.  As in the definition of 
$H_1(a)$, Eqs. (\ref{H1def},\, \ref{ieq1}), we should interpret the $i=1$ 
case of Eq. (\ref{completesum}) above as 
\BE
B(a) - \sum_{j=1}^{k} c_j a^j B_{j+1}(a) =1.
\EE
The proof parallels the steps above except that there is now a contribution 
from the $x=0$ endpoint that results in the cancellation of the $B(a)$ term.

\section*{Appendix B: Infrared limit and fixed-point behaviour}
\renewcommand{\theequation}{B.\arabic{equation}}
\setcounter{equation}{0}

     In asymptotically free theories, perturbation theory works best at 
high energies.  By investigating low energies we can learn important lessons 
about how perturbation theory can go bad, and why RS choice is crucial.  
As the physical scale $Q$ is lowered, and the effective couplant grows, 
we can expect the physical quantity ${\cal R}$ to either 
({\it i}) go to infinity at some finite energy of order $\tilde{\Lambda}$, 
or ({\it ii}) tend to a finite limit as $Q \to 0$.  The latter scenario is 
usually said to happen if and only if the $\beta$ function has a zero at some 
finite $a= a^*$, called a ``fixed point.''  That statement, however, is too 
naive because ``the $\beta$ function'' is not a unique object; it depends 
on RS.  

      At second order in QCD, for $n_f \! \le \! 8$, the $\beta$ function 
has no non-zero fixed point in any RS.  At higher orders, though, the 
question depends entirely on the RS choice.  In the $\msbar$ scheme the 
$c_2$ and $c_3$ coefficients are positive, so no fixed point exists and 
$\msbar$ results, at third and fourth orders, go to infinity at some $Q$ 
of order $\Lambda_{\smmsbar}$.  However, the optimization procedure does 
give finite results at arbitrarily low energies in the $e^+e^-$ case.  
At third order \cite{lowen} ${\cal R}(e^+e^-)$ approaches a finite limit, 
$0.3 \pm 0.3$.  At fourth order we find $0.2 \pm 0.1$.  These results can 
be found by applying the algorithm of Section \ref{sec:alg} at lower and 
lower $Q$, but can also be obtained much more simply because \cite{KSS} 
the optimization equations greatly simplify at a fixed point.  
The results of Ref.~\cite{KSS} (converting notation: 
$\rho_2^{\rm old} = \tilde{\rho}_2 - \quarter c^2$ and 
$\rho_3^{\rm old} = \half \tilde{\rho}_3$) are as follows.  
At third order the optimized $a^*$ is given by the smallest root of 
the quadratic equation
\BE 
\label{astar2}
\frac{7}{4} + c a^* + 
3 \left( \tilde{\rho}_2 -\frac{1}{4} c^2 \right) {a^*}^2 =0,
\EE
and the limiting value of ${\cal R}^{(3)}$ is 
\BE
{{\cal R}^*}^{(3)} = a^* \left( \frac{7}{6} + \frac{1}{6} c a^* \right).
\EE
At fourth order the corresponding equations are 
\BE 
\label{astar3}
\frac{83}{64} + \frac{13}{16} c a^* + \frac{3}{4} \tilde{\rho}_2 {a^*}^2 
+ 2 \tilde{\rho}_3 {a^*}^3 =0,
\EE
and 
\BE
{{\cal R}^*}^{(4)} = a^* \left( \frac{249}{256} + \frac{13}{64} c a^* + 
\frac{1}{16} \tilde{\rho}_2 {a^*}^2 \right).
\EE

     The fixed-point limit of Eq.~(\ref{sform}), the new formula for the 
$s_m \equiv \left( \frac{m+\Pp}{\Pp}\right) r_m$ coefficients, can be 
written as 
\BE
\hat{s}_m = \frac{1}{(k-1)} \left[ (k-2m-1) \hat{t}_m - 
(k-2m) \hat{t}_{m-1} \right],
\EE
or, equivalently,
\BE
\hat{s}_m = \frac{1}{(k-1)} \left[ (k-2m) \hat{c}_m - \hat{t}_m \right],
\EE
where $\hat{s}_m \equiv s_m {a^*}^m$, and $\hat{c}_m \equiv c_m {a^*}^m$, 
and $\hat{t}_m$ is a partial sum of $\beta$-function terms:  
\BE
\hat{t}_m = \sum_{i=0}^{m} c_j {a^*}^j.
\EE
These new formulas simplify the task of generalizing Ref.~\cite{KSS}'s 
results to higher orders.  

    The occurrence of a fixed point is not inevitable in OPT; the $a^*$ 
equation, (\ref{astar2}) or (\ref{astar3}), may or may not have a positive, 
real root.  A small positive $a^*$ is found for ${\cal R}(e^+e^-)$, for 
$n_f=2$, because the invariants $\tilde{\rho}_2$ and $\tilde{\rho}_3$ are 
negative and sizeable.  Interestingly, for $n_f$ above about $6$ (depending 
on the assumed electric charges of the extra quarks) one does not find a 
solution to Eq. (\ref{astar3}).  However, a finite infrared limit still 
exists, but it occurs by a new mechanism; we hope to report on this 
phenomenon in a future publication.

\end{document}